\documentclass[a4paper,nofootinbib,twocolumn]{revtex4}

\usepackage{graphicx}

\begin{document}

\title{Limits of Eliashberg Theory and Bounds for Superconducting
Transition Temperature}

\author{M.V. Sadovskii}

\affiliation{Institute for Electrophysics, Russian Academy of Sciences,
Ural Branch, Ekaterinburg 620016, Russia\\
E-mail: sadovski@iep.uran.ru}


\begin{abstract}

The discovery of record -- breaking values of superconducting transition 
temperature $T_c$ in quite a number of hydrides under high pressure was an
impressive demonstration of capabilities of electron -- phonon mechanism of
Cooper pairing. This lead to an increased interest to foundations and limitations
of Eliashberg -- McMillan theory as the main theory describing superconductivity
in a system of electrons and phonons. Below we shall consider both elementary
basics of this theory and a number of new results derived only recently.
We shall discuss limitations on the value of the coupling constant related to
lattice instability and a phase transition to another phase (CDW, bipolarons).
Within the stable metallic phase the effective pairing constant may acquire
arbitrary values. We consider extensions beyond the traditional adiabatic
approximation. It is shown that Eliasberg -- McMillan theory is also
applicable in the strong antiadiabatic limit. The limit of very strong
coupling, being most relevant for the physics of hydrides, is analyzed in details.
We also discuss the bounds for $T_c$ appearing in this limit.

\end{abstract}

\pacs{71.10.Fd, 74.20.-z, 74.20.Mn}

\maketitle


\tableofcontents


\section{Introduction}

The discovery \cite{H3S} of superconductivity with critical temperature approaching
$T_c=$ 203 K under pressure in the interval of 100-250 GPa (in diamond anvils)
in H$_3$S system lead to a stream of papers devoted to experimental studies
of high -- temperature superconductivity in hydrides in megabar region
(cf. reviews \cite{Er,ErD,ErPR}). Theoretical analysis immediately confirmed, that
these record -- breaking values of $T_c$ are ensured by traditional electron -- phonon
interaction in the limit of strong enough electron -- phonon coupling \cite{Ash,Grk-Krs}. 
More so, the detailed calculations for a number of rare earth hydrides under
pressure \cite{Ash} lead to the prediction of pretty large number of such systems
with record values of $T_c$. In some cases these predictions were spectacularly
confirmed. In particular experimentally the values of $T_c=$ 220 - 260 K were
obtained in systems like  LaH$_{10}$ \cite{DrEr,Som}, YH$_{6}$ \cite{P2},
(La,Y)H$_{6,10}$ \cite{P3}, YH$_{9}$ \cite{Snider}.
At last, quite recently a psychological border was crossed, when in Ref. \cite{RT} 
superconductivity with $T_c=$ 287.7$\pm$1.2 K (i.e. about +15 degrees Celsius) 
was obtained in C-H-S systems at pressures of 267$\pm$10 GPa.

The matter of principle here is that these works explicitly demonstrated the
absence of any significant limitations for $T_c$ within electron -- phonon
mechanism of Cooper pairing, where it was traditionally believed that
$T_c$ can not exceed 30-40 K. Correspondingly, the most pressing now became
the question of the upper limit of $T_c$, which can be achieved due to this
pairing mechanism.

Since the appearance of BCS theory it became obvious, that the increase of $T_c$
in superconductors can be realized by increasing the frequency of phonons
responsible for Cooper pairing, as well as by the increase of the effective
interaction of these phonons with electrons. These questions were studied by
numerous authors. The most developed approach for description of superconductivity
in the system of electrons of phonons remains Eliashberg -- McMillan theory
\cite{Grk-Krs,Scal,All,Kres,VIK}. It is well known that this theory is entirely
based upon the adiabatic approximation and Migdal theorem \cite{Mig,AGD,Schr,Diagr}, 
which allows to neglect vertex corrections while calculating the effects of
electron -- phonon interaction in typical metals. The real small parameter of 
perturbation theory here is $\lambda\frac{\Omega_0}{E_F}\ll 1$, where $\lambda$ 
is dimensionless constant of electron -- phonon interaction, $\Omega_0$ is 
characteristic phonon frequency and $E_F$ is Fermi energy of electrons. 
In particular, this leads to the conclusion that vertex corrections in this theory 
can be neglected even in case of $\lambda > 1$, because of inequality 
$\frac{\Omega_0}{E_F}\ll 1$ being valid in typical metals.
Recently, a number of papers appeared \cite{Est_1,Est_2,Est_3}, where some
doubts were expressed on these conclusions and some revisions proposed, 
based on the results of quantum Monte -- Carlo calculations for electron --
phonon system.

In Refs. \cite{MS_Eli,MS_Elis,MS_Elias} we have shown that under conditions of
strong nonadiabaticity, when  $\Omega_0\gg E_F$, the theory acquires 
new small parameter
$\lambda_D\sim \lambda\frac{E_F}{\Omega_0}\sim\lambda\frac{D}{\Omega_0}
\ll 1$ ($D$ is the half -- width of electron band), so that corrections to
electronic spectrum becomes insignificant. Vertex corrections can also be
neglected in this case, as was earlier shown in Ref. \cite{Ikeda}.
In general case the renormalization of electron spectrum (effective mass of
an electron) is determined by the new dimensionless coupling constant
$\tilde\lambda$, which tends to the usual $\lambda$ in adiabatic limit,
while in the strong antiadiabatic limit it tends to $\lambda_D$. 
At the same time, the temperature of superconducting transition $T_c$ 
in antiadiabatic limit is determined by the usual pairing constant of
Eliashberg -- McMillan theory $\lambda$, generalized for the account of the
finiteness of phonon frequency. Thus, the Eliashberg -- McMillan approach
remains valid also in the strong antiadiabatic limit.

In general, the interest towards the problem of superconductivity in the strong
antiadiabatic limit is stimulated by the discovery of a number of other
superconductors, where adiabatic approximation becomes invalid and characteristic
phonon frequencies are of the order of or even higher than electron Fermi energy.
Typical in this respect are intercalated systems with FeSe monolayers, as well 
as single -- layer films of FeSe on Sr(Ba)TiO$_3$ (FeSe/STO) substrates \cite{UFN}. 
The nonadiabatic nature of superconductivity in FeSe/STO system was first noted
by Gor'kov \cite{Gork_1,Gork_2}, while discussing the possible mechanism of
superconducting $T_c$ enhancement in FeSe/STO due to interaction with high -- energy
optical phonons of SrTiO$_3$ \cite{UFN}. Similar situation in fact appears also in
an old problem of superconductivity in doped SrTiO$_3$ \cite{Gork_3}, as well as in
twisted bi(tri)layers of graphene \cite{Graf}. In hydrides there are also 
possibility of existence of some small ``pockets'' of the Fermi surface with 
small values of Fermi energy \cite{Grk-Krs}. 

This paper is devoted to the critical review of these problems on rather
elementary level. Our presentation does not pretend to be exhaustive or giving
the complete review of multiple papers devoted to to studies of Eliashberg 
equations in recent decades. However, the author hopes that such presentation
can be useful both to young theorists and also to some experts in this field.

\section{Eliashberg -- McMillan approximation}

Fr\"ohlich Hamiltonian which is usually used to describe electron -- phonon
interaction can be written as \cite{Schr,Scal}:
\begin{eqnarray}
H=\sum_{\bf p}\varepsilon_{\bf p}a^+_{\bf p}a_{\bf p}
+\sum_{\bf k}\Omega_{0{\bf k}}b^+_{\bf k}b_{\bf k}+\nonumber\\
+\frac{1}{\sqrt N}\sum_{\bf pk}g_{\bf k}a^+_{\bf p+k}
a_{\bf p}(b_{\bf k}+b^+_{-\bf k})
\label{H_Frohlih}
\end{eqnarray}
where $\varepsilon_{\bf p}$ is electron spectrum counted from the Fermi level,
$\Omega_{0{\bf k}}$ is phonon spectrum\footnote{Note that here we have introduced
the ``bare'' phonon spectrum {\em in the absence} of electron -- phonon
interaction, which has no obvious definition in a real metal.}
and we have introduced the standard notations for creation $a^+_{\bf p}$ and
annihilation $a_{\bf p}$ operators of electrons and phonons --  $b^+_{\bf k}$ and
$b_{\bf k}$,  $N$ is the number of atoms in crystal.

The matrix element of electron -- phonon interaction has the following form
\cite{Schr,Scal}:
\begin{eqnarray}
g_{\bf k}=-\frac{1}{\sqrt{2M\Omega_{0{\bf k}}}}\langle{\bf p}|
{\bf e(\bf q)}\nabla V_{ei}({\bf r})|{\bf p+q}\rangle\nonumber\\
\equiv
-\frac{1}{\sqrt{2M\Omega_{0{\bf k}}}}I({\bf k})
\label{Fr_const}
\end{eqnarray}
where $V_{ei}$ is electron -- ion interaction potential, $M$ is ion mass,
and ${\bf e(q)}$ is polarization vector of a phonon with frequency $\Omega_{0\bf q}$.

To describe the phonon spectrum we often use simplified Debye and Einstein models.
In Debye model the phonon spectrum is assumed to be
$\Omega_{0{\bf k}}=ck$ ($c$ is the speed of sound) for all $k<k_D$, which gives an
elementary model of acoustic phonons. In this case Debye frequency $\Omega_D=ck_D$ 
defines the upper limit of phonon frequencies. In Einstein model phonon frequency is
assumed to be independent of the wave vector: $\Omega_{0{\bf k}}=\Omega_0$ for all
 $k$ within Brillouin zone, which gives the simplified model of optical phonons.

To describe interaction of electrons with optical (Einstein) phonons often the
so called Holstein model is also used. Its Hamiltonian is commonly written in
coordinate (site) representation in the lattice and electron -- phonon interaction
is assumed to be local (single  -- site):


\begin{eqnarray}
&&H=-t\sum_{ij\sigma} (t_{ij}a^+_{i\sigma}a_{j\sigma}+c.c.)+\Omega_0\sum_{i}
b^+_ib_i\ - \mu\sum_{i\sigma}n_{i\sigma} \nonumber\\
&&+g\sum_i(b_i+b^+_i)\sum_{\sigma}n_{i\sigma}
\label{Holst}
\end{eqnarray}
where $a^+_{i\sigma}$ and $a_{i\sigma}$ are creation and annihilation operators 
of electron with spin $\sigma$ on lattice site $i$, $n_i=a^+_{i\sigma}a_{i\sigma}$ is
electron density operator on site, $t_{ij}$ -- transfer integrals of electrons between
lattice sites, determining their spectrum (bandwidth) in tight -- binding approximation,
$b^+_i$ and $b_i$ corresponding operators for phonons with frequency $\Omega_0$.
The strength of electron -- phonon interaction is determined by interaction constant $g$.
Obviously this interaction describes the local interaction of Einstein phonon with
electron density at a lattice site. The chemical potential $\mu$ is determined by 
conduction band filling and defines the origin of the energy scale for electrons.

\begin{figure}
\includegraphics[clip=true,width=0.4\textwidth]{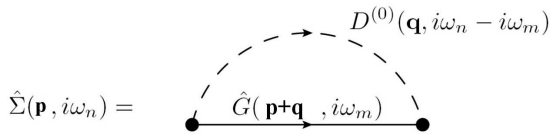}
\caption{Second -- order diagram for electronic self -- energy.}
\label{SE}
\end{figure}

Consider the simplest second -- order diagram of electron -- phonon interaction,
shown in Fig. \ref{SE}. Let us perform calculations in Matsubara technique
(i.e at finite temperatures $T>0$).
Analytical expression corresponding to this diagram has the form:
\begin{eqnarray}
\Sigma(i\omega_n, {\bf p})=-T\sum_{n=-\infty}^{\infty}\sum_{\bf p'}
|g_{\bf pp'}|^2G_0(i\omega_{n'},{\bf p'})\times\nonumber\\
\times D_0(i\omega_n-i\omega_{n'},
{\bf p-p'})
\label{SeFT2}
\end{eqnarray}
Subscript 0 at Green's functions of electron $G_0(i\omega_{n'},{\bf p'})$ 
and phonon $D_0(i\omega_n-i\omega_{n'},{\bf p-p'})$  (\ref{SeFT2})
indicates that these are Green's functions of free particles.

Summation over Matsubara frequencies is performed in a standard way
\cite{Schr,Diagr}, so that:
\begin{eqnarray}
\Sigma(i\omega_n,{\bf p})=\sum_{\bf p'}|g_{\bf pp'}
|^2\Biggl\{\frac{f_{\bf p'}+n_{\bf p-p'}}{i\omega_n - \varepsilon_{\bf p'}
+\Omega_{0\bf p-p'}}
\nonumber\\
+
\frac{1-f_{\bf p'}+n_{\bf p-p'}}
{i\omega_n-\varepsilon_{\bf p'}-\Omega_{0\bf p-p'}}\Biggr\}
\label{Self-Energy}
\end{eqnarray}
where $f({\bf p})=\frac{1}{e^{\frac{\varepsilon_{p}}{T}}+1}$ is Fermi distribition of
electrons, while
$n_{q}=\frac{1}{e^{\frac{\Omega_{0q}}{T}}+1}$ is Planckian (Bose) distribition of
phonons. For temperatures $T\to 0$ Fermi distribution of electrons transforms into
step -- function, while Planckian function of phonons tends to zero, so that the
first term in figure brackets is different from zero only for $\varepsilon_{p'}<0$,
while the second one for $\varepsilon_{p'}>0$.
Correspondingly, in the limit of $T=0$, after the substitution
$i\omega_n\to\varepsilon + i\delta sign\varepsilon_{p'}$, the contribution of
diagram of Fig. \ref{SE} can be written as \cite{Scal}:
\begin{eqnarray}
\Sigma(\varepsilon,{\bf p})=\sum_{\bf p'}|g_{\bf pp'}
|^2\Biggl\{\frac{f_{\bf p'}}{\varepsilon - \varepsilon_{\bf p'}
+\Omega_{0\bf p-p'}-i\delta}+\nonumber\\
+ \frac{1-f_{\bf p'}}
{\varepsilon-\varepsilon_{\bf p'}-\Omega_{0\bf p-p'} + i\delta}\Biggr\}
\label{self-energy}
\end{eqnarray}
Eq. (\ref{self-energy}) can be identically rewritten as:
\begin{eqnarray}
&&\Sigma(\varepsilon,{\bf p})=\nonumber\\
&&=\int d\omega\sum_{\bf p'}|g_{\bf pp'}|^2
\delta(\omega-\Omega_{0\bf p-p'})
\Biggl\{\frac{f_{\bf p'}}
{\varepsilon - \varepsilon_{\bf p'}+\omega-i\delta}+\nonumber\\
&&+\frac{1-f_{\bf p'}}
{\varepsilon - \varepsilon_{\bf p'}-\omega + i\delta}\Biggr\}
\label{self-energy_1}
\end{eqnarray}
Scattering of electrons by phonons in fact takes place in some narrow energy layer
close to Fermi level with the width of the order of double Debye frequency
 $2\Omega_D$, and in typical metals we always have $\Omega_D\ll E_F$. 
In this situation with high accuracy we can assume that both initial and final
momenta of electron ${\bf p}$ and ${\bf p'}$ are at the Fermi surface.
The main idea of Eliashberg -- McMillan approach is that we can avoid explicit
dependence on momenta, performing the averaging of the matrix element of
electron -- phonon interaction over surfaces of constant energy, corresponding to
initial and final momenta ${\bf p}$ and ${\bf p'}$, which is practically the same as
the averaging over corresponding real Fermi surfaces of a metal, which are
define by equations $\varepsilon({\bf p})=0$ and $\varepsilon({\bf p'})=0$. 
This is achieved by the following substitution ($N(0)$ is the density of states
at the Fermi level):
\begin{eqnarray}
|g_{\bf pp'}|^2\delta(\omega-\Omega_{0\bf p-p'})\Longrightarrow\nonumber\\
\frac{1}{N(0)}
\sum_{\bf p}\frac{1}{N(0)}\sum_{\bf p'}
|g_{\bf pp'}|^2\delta(\omega-\Omega_{0\bf p-p'})\delta(\varepsilon_{\bf p})
\delta(\varepsilon_{\bf p'})\nonumber\\
\equiv\frac{1}{N(0)}\alpha^2(\omega)F(\omega)
\label{Elias}
\end{eqnarray}
where in the final line we have introduced the standard {\em definition} of 
Eliashberg function $\alpha^2(\omega)$, which reflects the strength of electron -- phonon
interaction, while  $F(\omega)=\sum_{\bf q}\delta(\omega-\Omega_{0\bf q})$ is the
ohonon density of states, In principle, these functions can be directly determined
from experiments.

In the case when phonon energy becomes comparable or even exceeds Fermi energy,
electron scattering takes place not in a narrow layer close to  Fermi surface,
but in much wider energy interval. Then, for an initial
$|{\bf p}|\sim p_F$ the averaging over ${\bf p'}$ in expression like (\ref{Elias})
should be made over surface of constant energy, corresponding to
$E_F+\Omega_{0\bf p-p'}$ \cite{MS_Eli,MS_Elias}.
Correspondingly, Eq. (\ref{Elias}) is obviously generalized as:
\begin{eqnarray}
&&|g_{\bf pp'}|^2\delta(\omega-\Omega_{\bf p-p'})\Longrightarrow
\frac{1}{N(0)}
\sum_{\bf p}\frac{1}{N(0)}\sum_{\bf p'}\nonumber\\
&&|g_{\bf pp'}|^2
\delta(\omega-\Omega_{0\bf p-p'})\delta(\varepsilon_{\bf p})
\delta(\varepsilon_{\bf p'}-\Omega_{0\bf p-p'})\nonumber\\
&&\equiv\frac{1}{N(0)}\alpha^2(\omega)F(\omega)
\label{Elias_1}
\end{eqnarray}
which in the last $\delta$-function simply corresponds to a transition from
chemical potential $\mu$ to $\mu+\Omega_{\bf p-p'}$. Remember, that we always 
put the origin of an energy scale at $\mu=0$.

After the replacement (\ref{Elias}) the explicit dependence on momenta in
self -- energy vanishes and in the following we are dealing with
the average over the Fermi surface
$\Sigma(\varepsilon)\equiv\frac{1}{N(0)}
\sum_{\bf p}\delta(\varepsilon_{\bf p})\Sigma({\varepsilon,{\bf p}})$, which is
written now as:
\begin{eqnarray}
\Sigma(\varepsilon)=\int d\varepsilon'\int d\omega\alpha^2(\omega)F(\omega)
\Biggl\{\frac{f(\varepsilon')}
{\varepsilon - \varepsilon'+\omega-i\delta}+\nonumber\\
+ \frac{1-f(\varepsilon')}
{\varepsilon - \varepsilon'-\omega + i\delta}\Biggr\}
\label{self-energy_2}
\end{eqnarray}

In case of the self -- energy depending only on frequency (and not on momentum),
we can use the usual expressions for the (inverse) residue at the pole of Green's
function and electron mass renormalization \cite{Diagr}\footnote{Here we use
the notation inverse to that used in the theory of normal metals \cite{Diagr} 
to make it consistent with notations usually used in Eliashberg equations of
superconductivity theory. Correspondingly we have $Z\geq 1$, so that the residue
at the pole of Green's function is given by $Z^{-1}\leq 1$.}
\begin{equation}
Z=1-\left.\frac{\partial\Sigma(\varepsilon)}{\partial\varepsilon}
\right|_{\varepsilon=0}
\label{Z_res}
\end{equation}
\begin{equation}
m^{\star}=Zm=m\left.\left(1-\frac{\partial\Sigma(\varepsilon)}
{\partial\varepsilon}
\right|_{\varepsilon=0}\right)
\label{m_eff}
\end{equation}
Defining the dimensionless electron -- phonon coupling constant of Eliashberg --
McMillan theory as:
\begin{equation}
\lambda=2\int_{0}^{\infty}\frac{d\omega}{\omega}\alpha^2(\omega)F(\omega)
\label{lambda_Elias_Mc}
\end{equation}
by direct calculations we immediately obtain from (\ref{self-energy_2})
the standard expression for electron mass renormalization due to interaction
with phonons as:
\begin{equation}
m^{\star}=m(1+\lambda)
\label{mass_ren}
\end{equation}
The function $\alpha^2(\omega)F(\omega)$ in the expression for Eliashberg constant
of electron -- phonon interaction (\ref{lambda_Elias_Mc}) is to be calculated via
(\ref{Elias}) or is to be determined from experiments.

Using Eq. (\ref{Elias}) we can rewrite (\ref{lambda_Elias_Mc}) as:
\begin{equation}
\lambda=\frac{2}{N(0)}\int_{0}^{\infty}\frac{d\omega}{\omega}
\sum_{\bf p}\sum_{\bf p'}
|g_{\bf pp'}|^2
\delta(\omega-\Omega_{0\bf p-p'})\delta(\varepsilon_{\bf p})
\delta(\varepsilon_{\bf p'})
\label{Elias_lambda}
\end{equation}
which gives the standard way to calculate electron -- phonon coupling constant
$\lambda$, which determines, in particular, the Cooper pairing in Eliashberg -- McMillan
theory.

In the model of Einstein phonons $\Omega_{0k}\to \Omega_0$ and $g_{k}\to g_0$, 
so that dimensionless electron -- phonon coupling constant (\ref{Elias_lambda})
immediately reduces to the standard form \cite{Diagr}:
\begin{equation}
\lambda_0=\frac{2g_0^2 N(0)}{\Omega_0}
\label{dimep}
\end{equation} 

However, we must remember, that in general case the function
$\alpha^2(\omega)F(\omega)$ in the expression for Eliashberg electron -- phonon
coupling constant (\ref{lambda_Elias_Mc}) is to be calculated either via
(\ref{Elias}) or from (\ref{Elias_1}), depending on the ratio of Fermi energy
$E_F$ and characteristic phonon frequency $\Omega_0$.
Until $\Omega_0\ll E_F$ we can use the standard expression (\ref{Elias}),
while in the case of $\Omega)\sim E_F$ we have to use (\ref{Elias_1}).
Using Eq. (\ref{Elias_1}) we can rewrite (\ref{lambda_Elias_Mc}) in the following
form:
\begin{eqnarray}
\lambda=\frac{2}{N(0)}\int_{0}^{\infty}\frac{d\omega}{\omega}
\sum_{\bf p}\sum_{\bf p'}
|g_{\bf pp'}|^2\times\nonumber\\
\times\delta(\omega-\Omega_{0\bf p-p'})\delta(\varepsilon_{\bf p})
\delta(\varepsilon_{\bf p'}-\Omega_{0\bf p-p'})
\label{Elias_lambda_Om}
\end{eqnarray}
which gives the most general way to calculate electron -- phonon constant
$\lambda$, determining the Cooper pairing in Eliashberg -- McMillan theory.

\section{Migdal theorem}

Above, while calculating electron self -- energy due to electron -- phonon
interaction we have limited ourselves to a simplest contribution shown in
Fig. \ref{SE}. It may seem that we must take into account also other graphs
related to corrections to one of the vertices in this diagram. In fact this is
unnecessary, as these corrections to vertex part are small over adiabatic
parameter $\frac{\Omega_0}{E_F}\sim\sqrt{\frac{m}{M}}\ll 1$
(Migdal theorem) \cite{Mig} (cf. also Refs. \cite{AGD,Diagr,Schr,Scal}). 
Here $\Omega_0$ is characteristic phonon frequency of the order of Debye
frequency.

\begin{figure}
\includegraphics[clip=true,width=0.3\textwidth]{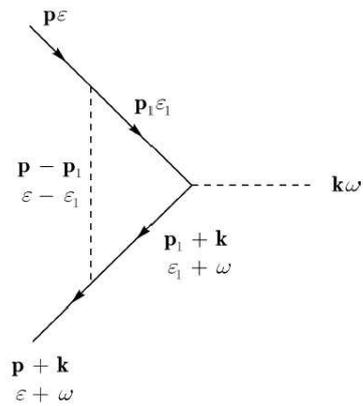}
\caption{Simplest vertex correction due to electron -- phonon interaction.}
\label{Mig_Th}
\end{figure}

Let us show this by estimating the simplest vertex correction determined by
graph shown in Fig. \ref{Mig_Th}.
We shall limit our analysis to a model with Einstein spectrum of phonons.
The analytic expression corresponding to diagram of Fig. \ref{Mig_Th} is:
\begin{eqnarray}
\Gamma^{(1)}=-g_0^3\int G_0({\bf p}_1\varepsilon_1)G_0({\bf p}_1+{\bf k},\varepsilon_1
+\omega)\times\nonumber\\
\times D_0(\varepsilon-\varepsilon_1,{\bf p-p}_1)
\frac{d^3p_1d\varepsilon_1}{(2\pi)^4}
\label{vxep}
\end{eqnarray}
Let us make a rough estimate of this expression. Consider first the integral
over $\varepsilon_1$. Assuming that the characteristic momentum transfer due
to phonon exchange is of the order of $k_D\sim p_F$, and taking into account
that $D_0(\varepsilon-\varepsilon_1)$ drops quadratically in the region of
$|\varepsilon-\varepsilon_1|\gg\Omega_0$, we obtain the main contribution
to this integral from the region of $|\varepsilon-\varepsilon_1|\sim\Omega_0$. 
Then the integral over $\varepsilon_1$ is of the order of 1, and we can write:
\begin{eqnarray}
\Gamma^{(1)}\sim
g_0^3\int d^3p_1\nonumber\\
\frac{1}{(\varepsilon_1-\varepsilon_{\bf p_1}+
i\delta sign\varepsilon_{\bf p_1})(\varepsilon_1+\omega-\varepsilon_{\bf p+k}
+i\delta sign\varepsilon_{\bf p_1+k})}\nonumber\\
\label{vxeph}
\end{eqnarray}
Consider now the remaining integral over $p_1$. Characteristic momentum transfer here
is of the order of $k_D\sim p_F$. Then we can estimate all denominators to be 
$\sim E_F$, and $\int d^3p_1\sim N(0)E_F$. Then we obtain:
\begin{equation}
\Gamma^{(1)}\sim g_0^3N(0)\frac{E_F}{E_F^2}\sim
g_0^3\frac{N(0)}{\Omega_0}\frac{\Omega_0}{E_F}\sim g_0\lambda_0\frac{\Omega_0}{E_F}
\label{estvx}
\end{equation}
Thus the relative size of this correction is:
\begin{equation}
\frac{\Gamma^{(1)}}{g_0}\sim\lambda_0\frac{\Omega_0}{E_F}
\sim\lambda_0\sqrt{\frac{m}{M}}
\label{Migd}
\end{equation}
where we have used $\frac{\Omega_0}{E_F}\sim\frac{\omega_D}{E_F}\sim\sqrt{\frac{m}{M}}$, 
where $m$ is electron mass, $M$ is ion mass. Electrons are much lighter than ions (nuclei)
and this correction to vertex part is negligible. More accurate analysis confirms this
conclusion \cite{AGD,Scal}, which is the essence of Migdal theorem.

Migdal's theorem allows us to neglect vertex corrections in calculations
related to electron -- phonon interaction in typical metals. The actual 
small parameter of perturbation theory is $\lambda_0\frac{\Omega_0}{E_F}\ll 1$, where
$\lambda_0$ is the dimensionless constant of electron -- phonon interaction,
$\Omega_0$ is characteristic phonon frequency, while $E_F$ is Fermi energy of
electrons, which in typical metals is of the order of conduction band width 
and determines the maximal energy scale. In particular this leads to a common 
belief, that vertex corrections in this theory can be neglected even in case of
$\lambda_0 > 1$, until inequality $\frac{\Omega_0}{E_F}\ll 1$ is valid, which is
characteristic for typical metals. In fact this means that taking into account  
the diagram of Fig. \ref{SE} only is sufficient even in the case of strong 
enough coupling between electrons and phonons.


Previous analysis implicitly assumed conduction band of infinite width.
In case of sufficiently large characteristic frequency of phonons it may become
comparable not only to Fermi energy, but also to conduction band width.
Curiously enough in the limit of very strong nonadiabaticity, when
$\Omega_0\gg E_F\sim D$ ($D$ is conduction band half -- width),  a new small 
parameter of perturbation theory $\lambda_D/\Omega_0\sim\lambda E_F/\Omega_0$
 
appears \cite{MS_Eli,MS_Elias}. Naturally, $\lambda$ in this case should be 
calculated using Eq. (\ref{Elias_lambda_Om}).

Consider the case of conduction band of finite width $2D$ with constant density
of states (two -- dimensional case). Fermi level as above is assumed to be at 
zero of energy scale and we assume the typical case of half -- filled band,
so that $E_F=D$.
Then Eq.  (\ref{self-energy_2}) reduces to:
\begin{eqnarray}
\noindent
\Sigma(\varepsilon)=\int_{-D}^{D} d\varepsilon'\int_{0}^{\infty} d\omega\alpha^2(\omega)F(\omega)
\Biggl\{\frac{f(\varepsilon')}
{\varepsilon - \varepsilon'+\omega-i\delta}+
\nonumber\\
+ \frac{1-f(\varepsilon')}
{\varepsilon - \varepsilon'-\omega + i\delta}\Biggr\}=\nonumber\\
=\int_{0}^{\infty} d\omega\alpha^2(\omega)F(\omega)
\Biggl\{\ln\frac{\varepsilon+D+\omega-i\delta}{\varepsilon-D-\omega+i\delta}
\nonumber\\
\noindent
-\ln\frac{\varepsilon+\omega-i\delta}{\varepsilon-\omega+i\delta}\Biggr\}
\label{A1}
\nonumber\\
\end{eqnarray}
Correspondingly, from Eq. (\ref{A1}) we get:
\begin{eqnarray}
-\left.\frac{\partial\Sigma(\varepsilon)}{\partial\varepsilon}
\right|_{\varepsilon=0}
=2\int_{0}^{D}d\varepsilon'\int_{0}^{\infty}d\omega\alpha^2(\omega)F(\omega)
\frac{1}{(\omega+\varepsilon')^2}
\nonumber\\
\noindent
=2\int_{0}^{\infty}d\omega\alpha^2(\omega)F(\omega)\frac{D}{\omega(\omega+D)}
\nonumber\\
\label{A5}
\end{eqnarray}
so that we can define the generalized coupling constant as:
\begin{equation}
\tilde\lambda=2\int_{0}^{\infty}\frac{d\omega}{\omega}\alpha^2(\omega)
F(\omega)\frac{D}{\omega+D}
\label{A6}
\end{equation}
which for $D\to\infty$ reduces to the usual Eliashberg -- McMillan constant
(\ref{lambda_Elias_Mc}), while for $D\to 0$ it gives the ``antiadiabatic'' coupling
constant:
\begin{equation}
\lambda_D=
2D\int \frac{d\omega}{\omega^2}\alpha^2(\omega)F(\omega)
\label{derivata_b}
\end{equation}
Eq. (\ref{A6}) describes a smooth crossover between the limits of wide and narrow
conduction bands.
Mass renormalization, in general, is determined by constant $\tilde\lambda$:
\begin{equation}
m^{\star}=m(1+\tilde\lambda)
\label{mass_renrm}
\end{equation}
For the model of one Einstein phonon with frequency $\Omega_0$ we have
$F(\omega)=\delta(\omega-\Omega_0)$, so that:
\begin{equation}
\tilde\lambda=\frac{2}{\Omega_0}\alpha^2(\Omega_0)\frac{D}{\Omega_0+D}
=\lambda\frac{D}{\Omega_0+D}= \lambda_D\frac{\Omega_0}{\Omega_0+D}
\label{A7}
\end{equation}
where Eliashberg -- McMillan coupling constant:
\begin{equation}
\lambda=2\int_{0}^{\infty}\frac{d\omega}{\omega}\alpha^2(\omega)F(\omega)=
\alpha^2(\Omega_0)\frac{2}{\Omega_0}
\label{lambda_Elias_Mc_opt}
\end{equation}
Comparison with Eq. (\ref{dimep}) gives $\alpha^2(\Omega_0)=g_0^2N(0)$ and
$\lambda_D$ reduces to:
\begin{equation}
\lambda_D=2\alpha^2(\Omega_0)\frac{D}{\Omega_0^2}=2\alpha^2(\Omega_0)
\frac{1}{\Omega_0}\frac{D}{\Omega_0}
\label{lamb_D}
\end{equation}
where in the last term we have explicitly written the new small parameter
$D/\Omega_0\ll 1$, appearing in the strong
antiadiabatic limit. Correspondingly, in this limit we always have:
\begin{equation}
\lambda_D=\lambda\frac{D}{\Omega_0}\sim\lambda\frac{E_F}{\Omega_0}\ll\lambda
\label{lamb_D_Mc}
\end{equation}
so that for reasonable values of $\lambda$ (up to the strong coupling region of
$\lambda\sim 1$) the  ``antiadiabatic'' coupling constant remains small.

It is obvious that vertex corrections also become small in this limit
as was shown by direct calculations in Ref. \cite{Ikeda}.
Thus we came to an unexpected conclusion --- in strong antiadiabatic limit 
electron -- phonon coupling becomes weak again! In this sense we can again speak of
validity of Migadl theorem also in antiadiabatic limit.
The physics here is simple -- in strong nonadiabatic limit ions move much faster
than electrons, so that electrons can not ``adjust'' to rapidly changing ion 
configurations and, in this sense, only are only weakly reacting to ion movements.


\section{Strong coupling and lattice instability}

\begin{figure}
\includegraphics[clip=true,width=0.45\textwidth]{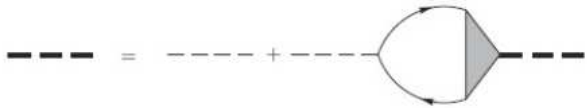}
\caption{Dyson equation for the full (``dressed'') phonon Green's function.}
\label{Phon_Dress}
\end{figure}

The general expression for phonon Green's function, taking into account the interaction
with electrons, is given by Dyson equation shown in Fig. {\ref{Phon_Dress}}. 
In analytic form we have:
\begin{equation}
D^{-1}({\bf k},i\omega_m)=D_{0}^{-1}({\bf k},i\omega_m)-|g_{\bf k}|^2\Pi({\bf k},i\omega_m)
\label{Phn_Green}
\end{equation}
Then we obtain:
\begin{equation}
D({\bf k},i\omega_m)=\frac{2\Omega_{0{\bf k}}}{(i\omega_m)^2-
\Omega_{0{\bf k}}^2-2\Omega_{0{\bf k}}|g_{\bf k}|^2\Pi({\bf k},i\omega_m)}
\label{PhGre}
\end{equation}
After the usual transition to real frequencies the phonon spectrum (renormalized by
interaction) is determined from the equation:
\begin{equation}
\Omega^2_{\bf k}=\Omega_{0{\bf k}}^2\left[1+\frac{2|g_{\bf k}|^2}{\Omega_{0{\bf k}}}
\Pi({\bf k},\Omega_{\bf k})\right]
\label{ph_spectr}
\end{equation}
In adiabatic approximation, taking into account Migdal theorem, polarization operator 
here can be taken as a simple loop. In the simplest case of free electrons we
have \cite{Diagr}:
\begin{equation}
\Pi({\bf k},\omega_m)=-2N(0)\left\{1+\frac{i\omega_m}{2v_Fk}
\ln\frac{i\omega_m-v_Fk}{i\omega_m+v_Fk}\right\}
\label{PolT00}
\end{equation}
or, after $i\omega_m\to\omega+i\delta$
\begin{equation}
\Pi({\bf k},\omega+i\delta )=-2N(0)\left\{1+\frac{\omega}{2v_Fk}\ln
\frac{\omega-v_Fk+i\delta}{\omega+v_Fk+i\delta}\right\}
\label{chiomq}
\end{equation}
where $v_F$ is electron velocity at Fermi surface, which much exceeds
the sound velocity, so that the values of $v_Fk$ are much higher than frequencies
of acoustical phonons, and for typical $k\sim p_F$ also those of optical phonons.
This again demonstrates the importance of adiabatic approximation in metals.
Thus, in calculations of phonon spectrum using (\ref{ph_spectr}) we can,
with high accuracy, immediately put $\omega=0$ in polarization operator.
In this case the imaginary part of polarization operator becomes zero and we simply
have $\Pi(0,0)=-2N(0)$. Then the phonon spectrum, renormalized by interaction with
electrons, is determined from:
\begin{equation}
\Omega^2_{\bf k}=\Omega_{0{\bf k}}^2\left[1+\frac{2|g_{\bf k}|^2}{\Omega_{0\bf k}}
\Pi(0,0)\right]=\Omega_{0\bf k}^2\left[1+\frac{\lambda_0^k}{N(0)}
\Pi(0,0)\right]
\label{ph_spectr_opt}
\end{equation}
and takes the form:
\begin{equation}
\Omega^2_{\bf k}=
\Omega_{0\bf k}^2[1-2\lambda_0^k]
\label{ph_spectr_inst}
\end{equation}
where we have introduced the usual definition of dimensionless coupling constant of
electron -- phonon interaction \cite{Diagr}:
\begin{equation}
\lambda_0^k=\frac{2|g_{\bf k}|^2 N(0)}{\Omega_{0{\bf k}}}
\label{dimepp}
\end{equation}
In this (rough enough) approximation the relatively small damping of phonons due
to electron -- phonon interaction just vanishes. It can be obtained with more
accurate treatment of the imaginary part of polarization operator
\cite{Diagr}.

The ``bare'' Green' function of phonons at real frequencies ($T=0$)
\begin{equation}
D_0({\bf k}\omega)=\frac{1}{\omega-\Omega_{0{\bf k}}+i\delta}-
\frac{1}{\omega+\Omega_{0{\bf k}}-i\delta}=\frac{2\Omega_{0{\bf k}}}
{\omega^2-\Omega^2_{0{\bf k}}+i\delta}
\label{fonGr}
\end{equation}
after such ``dressing'' by interaction with electrons transforms into
\cite{Diagr}:
\begin{equation}
D({\bf k}\omega)=\frac{2\Omega_{0{\bf k}}}
{\omega^2-\Omega^2_{\bf k}+i\delta}
\label{fonGrdress}
\end{equation}
where the renormalized phonon spectrum is given by (\ref{ph_spectr_inst}).

The spectrum given by Eq. (\ref{ph_spectr_inst}) signifies the lattice instability
for $\lambda_0^k>1/2$. This instability is often considered to be unphysical,
as was noted already in an early paper by Fr\"ohlich
\cite{Frol}, where it was obtained for the first time.
This point can be explained as follows. Let us rewrite the ``dressed''
Green's function (\ref{fonGrdress}) identically as:
\begin{equation}
D({\bf k}\omega)=\frac{2\Omega_{\bf k}}
{\omega^2-\Omega^2_{\bf k}+i\delta}\frac{\Omega_{0{\bf k}}}{\Omega_{\bf k}}
\label{fonGrdressed}
\end{equation}
Then it becomes clear that during diagram calculations, and Fig. \ref{SE} in particular
for electron self -- energy, using from the very beginning this renormalized
Green's function of phonons, the physical coupling constant of electron -- phonon
coupling takes the form (instead of (\ref{dimepp})):
\begin{equation}
\lambda^k=\frac{2|g_{\bf k}|^2 N(0)}{\Omega_{\bf{k}}}
\frac{\Omega_{0{\bf k}}}{\Omega_{\bf k}}=
\frac{2|g_{\bf k}|^2 N(0)}{\Omega_{0{\bf k}}}\frac{\Omega^2_{0{\bf k}}}{\Omega^2_{\bf k}}
=\lambda_0^k\frac{\Omega^2_{\bf k}}{\Omega^2_{\bf k}}
\label{dimepsc}
\end{equation}
or, using (\ref{ph_spectr_inst}):
\begin{equation}
\lambda^k=\frac{\lambda_0^k}{1-2\lambda_0^k}
\label{dimepscons}
\end{equation}
We see, that for $\lambda_0^k\to 1/2$ the {\em renormalized} coupling constant $\lambda^k$
monotonously grows and finally diverges. It is this costant that determines the ``true''
value of electron -- phonon interaction (with ``dressed'' phonons) and there is no
limitations for its value at all. This physical picture was discussed in detail, e.g.
in the famous book \cite{MaxKh}.

In a model with single Einstein phonon, which is a reasonable approximation for an optical
phonon, we have $\Omega_{\bf k}=\Omega_0$ and we can forget about dependence of
the coupling constant on phonon momentum, so that:
\begin{equation}
\lambda_0=\frac{2g_0^2 N(0)}{\Omega_0}
\label{dimep_ein}
\end{equation}
\begin{equation}
\Omega^2=\Omega_0^2[1-2\lambda_0]
\label{phn_spc_ein}
\end{equation}
\begin{equation}
\lambda=\frac{2g_0^2 N(0)}{\Omega_0}
\left(\frac{\Omega_0}{\Omega}\right)^2=\frac{\lambda_0}{1-2\lambda_0}
\label{dimepsc_ein}
\end{equation}
Eq. (\ref{dimepscons}) can be reversed and we can write:
\begin{equation}
\lambda_0^{k}=\frac{\lambda^k}{1+2\lambda^k}
\label{gbare}
\end{equation}
expressing nonphysical ``bare'' constant of electron -- phonon coupling
$\lambda_0^k$ via the ``true'' physical coupling constant $\lambda^k$.
Using this relation in the equation for renormalized phonon spectrum
(\ref{ph_spectr_inst}), we can write it as:
\begin{equation}
\Omega^{2}_{\bf k}=\Omega_{0{\bf k}}^2\left[1-\frac{2\lambda^k}{1+2\lambda^k}\right]
=\Omega_{0{\bf k}}^2\frac{1}{1+2\lambda^k}
\label{phn_spc_dressed}
\end{equation}
so that in this representation there is no instability of spectrum (lattice), and the
growth of $\lambda^k$ just leads to continuous ``softening'' of spectrum due to the
growth of electron -- phonon coupling.

In a model of Einstein phonon all relations simplify and we get:
\begin{equation}
\lambda_0=\frac{\lambda}{1+2\lambda}
\label{gbare_eins}
\end{equation}
\begin{equation}
\Omega=\frac{\Omega_0}{\sqrt{1+2\lambda}}
\label{phn_spc_dressed_eins}
\end{equation}

In Eliashberg -- McMillan formalism, where we perform the averaging over the momenta of
electrons on the (arbitrary) Fermi surface, McMillan function $\alpha^2(\omega)F(\omega)$, 
naturally should be determined bu the physical (renormalized) spectrum of phonons:
\begin{equation}
\alpha^2(\omega)F(\omega)=\frac{1}{N(0)}\sum_{\bf pp'}|g_{\bf pp'}|^2
\delta(\omega-\Omega_{\bf p-p'})\delta(\varepsilon_{\bf p})
\delta(\varepsilon_{\bf p'})
\label{McM_El}
\end{equation}
In particular case of Einstein phonon it immediately reduces to (\ref{dimepsc_ein})
and there is no limitations on the value of $\lambda$.

In self -- consistent derivation of Eliashberg equations we have to use the
diagram of Fig. \ref{SE}, where the the phonon Green's function is taken in ``dressed'' 
form (\ref{fonGrdress}) or (\ref{fonGrdressed}) and describes the physical
(renormalized) phonon spectrum. In this case we {\em do not have to} include corrections 
to this function due to electron -- phonon interaction, as they are already taken into account
in phonon spectrum (\ref{ph_spectr_inst}).

It should be noted that the value of critical coupling constant obtained above, at which 
Fr\"ohlich instability of phonon spectrum appears, is obviously directly related to the use
of the simplest expression for polarization operator of the gas of free electrons
(\ref{PolT00}), (\ref{chiomq}), which was calculated neglecting vertex corrections and
self -- consistent ``dressing'' of electron Green's functions entering the loop.
Naturally, even in the simplest cases like the problem with Einstein spectrum accounting
for these higher corrections, as well as more realistic structure of electron spectrum
in a lattice, can somehow change the value of  $\lambda_0$, corresponding to instability
of the ``bare'' phonon spectrum, so that it will differ from 1/2. In this sense it is
better to speak about instability at some ``critical'' value  $\lambda_0^{c}\sim 1/2$.

In general case the inverse influence of electrons on phonons can be taken into account by
generalizing Eq. (\ref{McM_El}) in the following way:
\begin{eqnarray}
&&\alpha^2(\omega)F(\omega)=\frac{1}{N(0)}\sum_{\bf pp'}|g_{\bf pp'}|^2
\left(-\frac{1}{\pi}Im D^R(\omega,{\bf p-p'})\right)\nonumber\\
&&\times\delta(\varepsilon_{\bf p})
\delta(\varepsilon_{\bf p'})=\nonumber\\
&&=\frac{1}{N(0)}\sum_{\bf pp'}|g_{\bf pp'}|^2
B({\bf p-p'},\omega)\delta(\varepsilon_{\bf p})
\delta(\varepsilon_{\bf p'})
\label{McM_El_D}
\end{eqnarray}
where we have introduced phonon spectral density $B({\bf q},\omega)$, which determines the phonon
Green's function (in Matsubara representation) via the spectral relation:
\begin{equation}
D({\bf q},i\omega_m)=\int_{0}^{\infty}B({\bf q},\omega)\frac{2\omega}{(i\omega_m^2)
-\omega^2}d\omega
\label{spD}
\end{equation}
In particular, from here we get:
\begin{equation}
D({\bf q},0)=-2\int_{0}^{\infty}\frac{d\omega}{\omega}B({\bf q},\omega)
\label{Dq0}
\end{equation}
Then we can write the following general relation for Eliashberg -- McMillan constant $\lambda$:
\begin{eqnarray}
&&\lambda=2\int_{0}^{\infty}\frac{d\omega}{\omega}\alpha^2(\omega)F(\omega)=
\nonumber\\
&&=\frac{2}{N(0)}\int_{0}^{\infty}\frac{d\omega}{\omega}\sum_{\bf pp'}|g_{\bf pp'}|^2
B({\bf p-p'},\omega)\delta(\varepsilon_{\bf p})
\delta(\varepsilon_{\bf p'})=\nonumber\\
&&=-\frac{1}{N(0)}\sum_{\bf pp'}|g_{\bf pp'}|^2
D({\bf p-p'},0)\delta(\varepsilon_{\bf p})
\delta(\varepsilon_{\bf p'})
\label{lamb_gen}
\end{eqnarray}
For the model of optical phonon with frequency $\Omega_0$ this immediately reduces to:
\begin{equation}
\lambda=-g_0^2N(0)\left\langle D({\bf p-p'},0)\right\rangle_{FS}
=-\frac{\lambda_0\Omega_0}{2}\left\langle D({\bf p-p'},0)\right\rangle_{FS}
\label{lambd_Est}
\end{equation}
where we have introduced the usual notation for momentum averaging over Fermi surface.

\begin{figure}
\includegraphics[clip=true,width=0.35\textwidth]{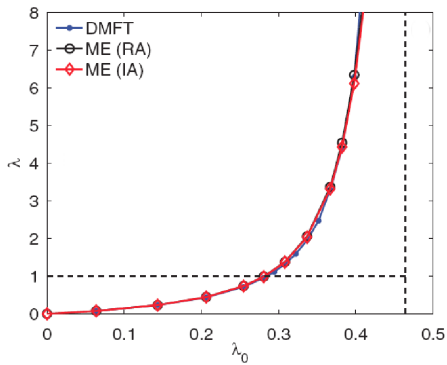}
\caption{Dependence of the renormalized coupling constant $\lambda$, calculated from
(\ref{lambd_Est}), on the ``bare'' $\lambda_0$, obtained within self -- consistent 
Eliashberg theory (ME(RA) -- real frequency technique, ME(IA) -- Matsubara technique)
and in DMFT(QMC) \cite{Gunn}. Vertical dashed line corresponds to 
$\lambda_{0}^{c}\approx$0.464.}
\label{DMFT-1}
\end{figure}

\begin{figure}
\includegraphics[clip=true,width=0.35\textwidth]{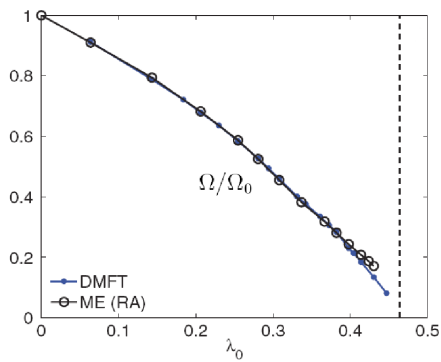}
\caption{Dependence of renormalized Einstein phonon frequency $\Omega$
on ``bare'' $\lambda_0$, obtained within self -- consistent 
Eliashberg theory (ME(RA) -- real frequency technique, ME(IA) -- Matsubara technique)
and in DMFT(QMC) \cite{Gunn}. Vertical dashed line corresponds to 
$\lambda_{0}^{c}\approx$0.464.}
\label{DMFT-2}
\end{figure}

The previous analysis can be significantly improved within simplified Holstein model
(\ref{Holst}), where the local (single site) nature of interaction allows solving
it using the dynamical mean field theory (DMFT) \cite{DMFT1,DMFT2,DMFT3,DMFT4}, 
which becomes (numerically) exact in the limit of lattice of infinite dimensions
(infinite number of nearest neighbors). Such analysis was performed e.g. in Ref. \cite{Gunn},
using as quantum Monte -- Carlo (QMC) as impurity solver in DMFT. The main results are
shown in Figs. \ref{DMFT-1}, \ref{DMFT-2}.

In particular in Fig. \ref{DMFT-1} we show the dependence of renormalized
$\lambda$ on ``bare'' $\lambda_0$. It can be seen, that the usual behavior of
Fr\"ohlich theory is nicely reproduced with slightly changed $\lambda_{0}^{c}$=0.464. 
Similar behavior is observed for renormalized phonon frequency $\Omega$, as seen in
Fig. \ref{DMFT-2}. Rather insignificant deviations from predictions of
Eliashberg theory are observed only in immediate vicinity of
$\lambda_{0}^{c}$, especially for $\lambda_0>$0.4. 

The instability appearing at $\lambda_0=\lambda_{0}^{c}$ in Holstein model 
(in DMFT approximation) with half -- filled bare band, was convincingly
interpreted in Ref. \cite{Bull} as transition into the state of {\em bipolaron
insulator}. Until this transition the system remains metallic and is nicely 
described by Eliashberg theory (with insignificant numerical corrections). 

In a series of papers \cite{Est_1, Est_2, Est_3} direct calculations by 
dynamical quantum Monte -- Carlo (DQMC) were performed for a number of 
characteristics of Holstein model on two -- dimensional (square) lattice. 
The results obtained were in some respects similar to the conclusions of
Refs. \cite{Gunn,Bull} --- up to the values of the ``bare'' constant 
$\lambda_0$=0.4 there is a good agreement with predictions of Eliashberg
theory, but in the interval of  $\lambda_0$ from 0.4 to 0.5
certain deviations are observed. This interval of $\lambda_0$ values, according
to the same calculations, corresponds to the interval of renormalized
$\lambda$ from 1.7 to 4.6. For $\lambda_0\approx$ 0.5 system undergoes transition
into the state of bipolaron insulator with commensurate charge density wave (CDW).

It is quite clear that the critical value of electron -- phonon coupling constant
obtained above, corresponding to Fr\"ohlich instability of phonon spectrum at
$\lambda_0$=1/2, was the direct result of our use of the simplest expression
for polarization operator of free electron gas (\ref{PolT00}), (\ref{chiomq}), 
where there was no significant dependence on the wave vector $\bf k$. 
Naturally, this dependence is also absent in DMFT approximation.
If we go to some more realistic model of electron spectrum, like tight -- binding 
approximation in some definite crystal lattice (not in infinite dimensions!), 
we can obtain instability of phonon spectrum at some finite value of
phonon wave vector $\bf k$ \cite{Est_1,Est_2,Est_3,Opp}. The appearance of such 
instabilities, as is well known, usually corresponds to formation in the system
of charge density waves (CDW) \cite{Diagr}. In the case of the ``nesting''
properties of Fermi surface, these instabilities appear (at $T=0$) even for 
infinitesimal values of the coupling constant of electron -- phonon interaction
\cite{Diagr}. After that, metal acquires the new ground state (of dielectric nature)
so that all theoretical analysis is to be performed in a new way.
In general case, considered here, this instability appears at finite (large enough)
values of the ``bare'' constant of electron -- phonon interaction.
Naturally, the usual Eliashberg -- McMillan theory ``works'' only within usual
metallic phase, which is of the main interest to theory of superconductivity.
In this sense, we are in certain disagreement with terminology of Refs.
\cite{Est_1, Est_2, Est_3}, where it was claimed, that Eliashberg theory becomes 
invalid for the values of $\lambda_0\sim\lambda_{0}^{c}$. In reality, in this rather
narrow region we are observing corrections due to closeness of the system to
instability --- the phase transition into a new ground state (bipolarons, CDW),
where become important fluctuations of corresponding order parameter.
Eliashberg theory, considered as a mean -- field theory, nicely describes almost
all metallic region, except this ``critical'' neighborhood of $\lambda_{0}^{c}$, 
including large enough values of {\em physical} coupling constant $\lambda$ 
(which simply diverges at this transition).

It should be stressed here, that all conclusions on instability of metallic
phase were done above in the framework of purely {\em model} approaches
(Fr\"ohlich and Holstein models) and in terms of ``bare'' parameters of
these models like $\lambda_0$ and $\Omega_0$, which, as was often noted in the
literature, are not so well defined physically. This is known for a long time and
was discussed many times. The problem nere is that the phonon spectrum in a metal,
considered as system of ions and electrons, is usually assumed to be calculated 
in adiabatic approximation \cite{BK}. This spectrum is relatively weakly
renormalized due to nonadiabatic effects, which are small over the small
parameter $\sqrt{\frac{m}{M}}$ \cite{BK,Geilik}.
In this respect, it is drastically different from the ``bare'' spectra of
Fr\"ohlich or Holstein models, which, as we have seen above, is significantly
renormalized by electron -- phonon interaction. The physical meaning of the
``bare'' spectrum $\Omega_0$ in these models remains not so clear, in contrast
to phonon spectrum in metals, calculated in adiabatic approximation.
In any case it can not be determined from any experiments, Similarly, the
same can be said about the parameters of electron -- phonon interaction.
There were numerous attempts to build a consistent theory of electron -- phonon
interaction on the background of physical (adiabatic) phonon spectrum \cite{Geilik}, 
but these were not so successful. Rather detailed discussion of the modern state
of this problem can be found in Ref. \cite{Maks}. The short recipe for practical
calculations is to identify the renormalized (``dressed'') spectrum of phonons
$\Omega$ in Fr\"ohlich or Holstein models with physical (adiabatic)
phonon spectrum, which is not to be further renormalized, and should be taken
from adiabatic calculations or from the experiment\footnote{Ideologically,
situation here is quite analogous to the standard approach in quantum electrodynamics,
where the physical charge and mass of an electron are defined by infinite series
of perturbation theory and are taken from the experiment.} Precisely this point of view is
usually implicitly accepted in calculations within Eliashberg -- McMillan theory.
Until the system remains in metallic phase, this point of view remains quite consistent.
Then there are practically no limitations on the value of physical (renormalized)
coupling constant $\lambda$ and Eliashberg -- McMillan theory remains valid up to
its large enough values (limited only by Migdal theorem).

\section{McMillan expression for electron -- phonon coupling constant}

McMillan has derived a simple expression for the dimensionless electron -- phonon
coupling in Eliashberg theory \cite{VIK}. Let us write down Eliashberg -- McMillan
function (\ref{Elias}) using (\ref{Fr_const}) as:
\begin{eqnarray}
&&\alpha^2(\omega)F(\omega)=\nonumber\\
&&=\frac{1}{N(0)}
\sum_{\bf p}\sum_{\bf p'}
|g_{\bf pp'}|^2\delta(\omega-\Omega_{\bf p-p'})\delta(\varepsilon_{\bf p})
\delta(\varepsilon_{\bf p'})=\nonumber\\
&&=\frac{1}{N(0)}\sum_{\bf p}\sum_{\bf p'}
\frac{1}{2M\Omega_{\bf p-p'}}|I({\bf p-p'})|^2\delta(\omega-\Omega_{\bf p-p'})\nonumber\\
&&\times\delta(\varepsilon_{\bf p})
\delta(\varepsilon_{\bf p'})
\label{Elias_McMil}
\end{eqnarray}
where $\Omega_{\bf p-p'}$ is assumed to be the physical frequency of phonons.
Then we immediately get:
\begin{equation}
\int_{0}^{\infty}d\omega\alpha(\omega)F(\omega)\omega=\frac{N(0)
\langle I^2\rangle}{2M}
\label{I2M}
\end{equation}
Now rewrite (\ref{lambda_Elias_Mc})as:
\begin{eqnarray}
\lambda=2\int_{0}^{\infty}\frac{d\omega}{\omega}\alpha^2(\omega)F(\omega)=
\nonumber\\
=\frac{1}{\langle\Omega^2\rangle}\int_{0}^{\infty}d\omega\alpha^2(\omega)F(\omega)\omega
\label{lambda_Eliashb_Mc}
\end{eqnarray}
where the mean square phonon frequency is defined as:
\begin{eqnarray}
\langle\Omega^2\rangle = 
\frac{\int_{0}^{\infty}d\omega\alpha^2(\omega)F(\omega)\omega}
{\int_{0}^{\infty}\frac{d\omega}{\omega}\alpha^2(\omega)F(\omega)}=
\nonumber\\
=\frac{2}{\lambda}\int_{0}^{\infty}d\omega
\alpha^2(\omega)F(\omega)\omega\nonumber\\
\label{aver_sq_w}
\end{eqnarray}
From this expression we can immediately see that:
\begin{equation}
\lambda=\frac{N(0)\langle I^2\rangle}{M\langle\Omega^2\rangle}
\label{McM_form}
\end{equation}
where we have introduces the matrix element of the gradient of electron --
ion potential averaged over Fermi surface:
\begin{eqnarray}
\langle I^2\rangle=\frac{1}{[N(0)]^2}\sum_{\bf p}\sum_{\bf p'}
\left|I({\bf p-p'})\right|^2\delta(\varepsilon_{\bf p})
\delta(\varepsilon_{\bf p'})=\nonumber\\
=\frac{1}{[N(0)]^2}\sum_{\bf p}\sum_{\bf p'}
\left|\langle{\bf p}|
\nabla V_{ei}({\bf r})|{\bf p'}\rangle\right|^2)\delta(\varepsilon_{\bf p})
\delta(\varepsilon_{\bf p'})=\nonumber\\
=\langle |\langle {\bf p}|\nabla V_{ei}({\bf r})|{\bf p'}\rangle|^2\rangle_{FS}
\label{grV2}
\end{eqnarray}
Eq. (\ref{McM_form}) gives very useful representation for $\lambda$, which is
often used in the literature and in practical calculations.

\section{Eliashberg equations}

Eliashberg theory is based on a system of equations for normal and anomalous 
Green's functions of a superconductor \cite{Diagr}.
Obviously, the solution of these {\em integral} equations, taking into account
the real spectrum of phonons, represents rather difficult problem. 
However, a significant progress was achieved and the theory of traditional
superconductors, based on the picture of pairing due to electron -- phonon
interaction, is an example of very successful application of Green's functions.
Very good presentation of methods used and applications of Eliashberg equations
can be found in Ref. \cite{VIK}. 

Below we shall present somehow simplified derivation of Eliashberg equations,
dropping some technical details. In particular, we shall not consider the role
of direct Coulomb repulsion of electrons within Cooper pair, which is accounted
for in the complete Eliashberg -- McMillan theory \cite{VIK}, limiting ourselves
only to electron -- phonon interaction. The accounting for Coulomb contributions
is not especially difficult \cite{VIK} and reduces at the end to introduction of
Coulomb pseudopotential $\mu^{\star}$ \cite{VIK}, which in typical metals is 
rather small and not so important in the limit of very strong coupling with phonons,
which will be of the main interest for us in the following\footnote{Surely, accounting 
for $\mu^{\star}$ is very important for quantitative estimates of superconducting
transition temperature in the weak and intermediate coupling region.}

Taking into account Migdal theorem, in adiabatic approximation vertex corrections
are irrelevant, so that Eliashberg equations can be derived by calculating the
diagram of Fig. \ref{SE}, where electron Green's function in superconducting state
is taken in Nambu matrix representation \cite{Schr}. Calculations are similar to
derivation of (\ref{self-energy_2}) and can be performed in Matsubara technique
($T\neq 0$).  In Nambu formalism electronic Green's function of superconductor is
written in a standard way as ($\hat\sigma_i$ are Pauli matrices) \cite{VIK}:
\begin{equation}
\hat G^{-1}(i\omega_n,{\bf p})=i\omega_n\hat 1 - \varepsilon_{\bf p}\hat\sigma_z
-\hat\Sigma(i\omega_n,{\bf p})
\label{G_Nambu_Mat}
\end{equation}
where matrix self -- energy is represented as\footnote{Possible contribution
proportional to $\hat\sigma_y$ is removed by the appropriate choice of the
phase of superconducting order parameter, while the term proportional to
$\hat\sigma_z$ reduces to renormalization of chemical potential \cite{VIK}.}:
\begin{equation}
\hat \Sigma(i\omega_n,{\bf p})=(1-Z(i\omega_n))i\omega_n\hat 1+Z(i\omega_n)
\Delta(i\omega_n)\hat\sigma_x
\label{Sigma_Nambu_Mat}
\end{equation}
Here we are introducing a number of simplifying assumptions like independence
of renormalization factor $Z(i\omega_n)$ and gap function $\Delta(i\omega_n)$
on momentum \cite{VIK}. Then we have:
\begin{equation}
\hat G(i\omega_n,{\bf p})=\frac{Z(i\omega_n)i\omega_n\hat 1+\varepsilon_{\bf p}
\hat\sigma_z+Z(i\omega_n)\Delta(i\omega_n)\hat\sigma_x}{Z^2(i\omega_n)
(i\omega_n)^2-Z^2(i\omega_n)\Delta^2(i\omega_n)-\varepsilon^2_{\bf p}}
\label{G_Nambu_Mats}
\end{equation}
Self -- energy part corresponding to diagram like Fig. \ref{SE} with matrix Green's
function of electron (\ref{G_Nambu_Mats}) is written as:
\begin{eqnarray}
&&\hat\Sigma(\omega_n,{\bf p})=-T\sum_m\sum_{\bf p'}|g_{\bf pp'}|^2
D(i\omega_n-i\omega_m,{\bf p-p'})\times\nonumber\\
&&\times\hat\sigma_z\hat G({i\omega_m,\bf p'})\hat\sigma_z
\label{SE_Nambu_phon}
\end{eqnarray}
where phonon Green's function $D(i\omega_n-i\omega_m,{\bf p-p'})$ can be taken as in
Eq. (\ref{fonGr}), denoting the phonon frequency $\Omega_{\bf p-p'}$ as in Eq. (\ref{self-energy}).

As we know, all physics of conventional superconductivity develops in a layer with
the width of the order of $2\Omega_D\ll E_F$ close to Fermi surface.
Thus we can make here the substitution (\ref{Elias}) and obtain from (\ref{SE_Nambu_phon})  
the expression for self -- energy part averaged over momenta on Fermi surface
(similar to (\ref{self-energy_2})). 
As a result, we obtain the general system of equations for the gap $\Delta(\omega_n)$ 
and renormalization factor $Z(\omega_n)$ of the following form:
\begin{eqnarray}
&&\Delta(\omega_n)Z(\omega_n)=T\sum_{n'}\int_{-\infty}^{\infty}d\xi\int_{0}^{\infty}
d\omega\alpha^2(\omega)F(\omega)\times\nonumber\\
&&\times D(\omega_n-\omega_{n'};\omega)\frac{\Delta(\omega_n')}
{\omega^2{_n'}+\xi^2+\Delta^2(\omega_{n'})}
\label{El_Mats}
\end{eqnarray}
\begin{eqnarray}
&&1-Z(\omega_n)=\frac{\pi T}{\omega_n}\sum_{n'}\int_{-\infty}^{\infty}d\xi\int_{0}^{\infty}d\omega
\alpha^2(\omega)F(\omega)\times\nonumber\\
&&\times D(\omega_n-\omega_{n'};\omega)
\frac{\omega_n'}
{\omega^2{_n'}+\xi^2+\Delta^2(\omega_{n'})}
\label{Zgener}
\end{eqnarray}
where we have introduced
\begin{equation}
D(\omega_n-\omega_{n'};\omega)=\frac{2\omega}{(\omega_n-\omega_{n'})^2+\omega^2}
\label{Dw}
\end{equation}
The integral over $\xi$ here is easily calculated and gives:
\begin{eqnarray}
&&\int_{-\infty}^{\infty}d\xi\frac{1}{\omega_{n'}^2+\xi^2+\Delta^2(\omega_{n'})}=
\frac{\pi}{\sqrt{\omega_{n'}^2+\Delta^2(\omega_{n'})}}\to\nonumber\\
&&\to\frac{\pi}{|\omega_{n'}|}\ 
\mbox{for}\ \Delta(\omega_{n'})\to 0
\label{intxi}
\end{eqnarray}
Then the linearized gap equation (equation for $T_c$) has the form:
\begin{eqnarray}
\Delta(\omega_n)Z(\omega_n)=\pi T\sum_{n'}\int_{0}^{\infty}\alpha^2(\omega)
F(\omega)\times\nonumber\\
\times D(\omega_n-\omega_{n'};\omega)\frac{\Delta(\omega_{n'})}{|\omega_{n'}|}
\label{lin_delta}
\end{eqnarray}
where
\begin{eqnarray}
&&1-Z(\omega_n)=\frac{\pi T}{\omega_n}\sum_{n'}\int_{-\infty}^{\infty}d\xi\int_{0}^{\infty}d\omega
\alpha^2(\omega)F(\omega)\times\nonumber\\
&&\times D(\omega_n-\omega_{n'};\omega)
\frac{\omega_{n'}}
{|\omega{_n'}|}
\label{lin_Zgen}
\end{eqnarray}
The general gap equation is:
\begin{eqnarray}
&&\Delta(\omega_n)Z(\omega_n)=\pi T\sum_{n'}\int_{0}^{\infty}\alpha^2(\omega)
F(\omega)\times\nonumber\\
&&\times D(\omega_n-\omega_{n'};\omega)
\frac{\Delta(\omega_{n'})}
{\sqrt{\omega^2_{n'}+\Delta^2(\omega_{n'})}}
\label{eq_delta}
\end{eqnarray}
where factor $Z(\omega_n)$ is determined from the following equation:
\begin{eqnarray}
&&1-Z(\omega_n)=\frac{\pi T}{\omega_n}\sum_{n'}\int_{0}^{\infty}d\omega
\alpha^2(\omega)F(\omega)\times\nonumber\\
&&\times D(\omega_n-\omega_{n'};\omega)
\frac{\omega_{n'}}{\sqrt{\omega_{n'}^2+\Delta^2(\omega_{n'})}}
\label{Zgen}
\end{eqnarray}
to be solved jointly with (\ref{eq_delta}).

In a model with Einstein spectrum of phonons Eq. (\ref{eq_delta}) reduces to:
\begin{eqnarray}
\Delta(\omega_n)Z(\omega_n)=\pi T\lambda\sum_{n'}\frac{\Omega_0^2}{(\omega_n-
\omega_{n'})^2+\Omega_0^2}\times\nonumber\\
\times\frac{\Delta(\omega_{n'})}{\sqrt{\omega^2_{n'}+
\Delta^2(\omega_{n'})}}
\label{delta_gen_eins}
\end{eqnarray}
while  Eq. (\ref{Zgen}) becomes:
\begin{equation}
Z(\omega_n)=1+\frac{\pi T\lambda}{\omega_n}\sum_{n'}\frac{\Omega_0^2}{(\omega_n-
\omega_{n'})^2+\Omega_0^2}\frac{\omega_{n'}}{\sqrt{\omega_{n'}^2+\Delta^2(\omega_{n'})}}
\label{Z_eins_gen}
\end{equation}
where the coupling constant $\lambda$, determined by standard expressions 
(\ref{lambda_Elias_Mc}) or (\ref{lambda_Elias_Mc_opt}), appears explicitly.

To determine $T_c$ (in the limit of $\Delta(\omega_{n'})\to 0$) in a system with
Einstein spectrum of phonons we obtain the following system of linear homogeneous
Eliashberg equations: 
\begin{equation}
\Delta(\omega_n)Z(\omega_n)=\pi T\lambda\sum_{n'}\frac{\Omega^2_0}{(\omega_n-
\omega_{n'})^2+\Omega_0^2}\frac{\Delta(\omega_{n'})}{|\omega_{n'}|}
\label{lin_delta_einst}
\end{equation}
where
\begin{equation}
Z(\omega_n)=1+\frac{\pi T\lambda}{\omega_n}\sum_{n'}\frac{\Omega_0^2}{(\omega_n-
\omega_{n'})^2+\Omega_0^2}\frac{\omega_{n'}}{|\omega_{n'}|}
\label{Z_lin_einst}
\end{equation}
It is clear that the value of $T_c$ is determined by zero determinant of this system
of equations.

Note that in general equations (\ref{eq_delta}), (\ref{Zgen}) the coupling constant
$\lambda$ does not appear explicitly. Usually this is achieved by reduction of these
equations to ``Einstein'' form like (\ref{delta_gen_eins}), (\ref{Z_eins_gen}) by
introduction of the average square of phonon frequency, defined as:
\begin{equation}
\langle\Omega^2\rangle = \frac{2}{\lambda}\int_{0}^{\infty}d\omega
\alpha^2(\omega)F(\omega)\omega
\label{av_sq_w}
\end{equation}
and the following replacement in (\ref{eq_delta}), (\ref{Zgen}):
\begin{equation}
\frac{1}{(\omega_n-\omega_{n'})^2+\omega^2}
\Longrightarrow \frac{1}{(\omega_n-\omega_{n'})^2
+\langle\Omega^2\rangle}
\label{Dw_eff}
\end{equation}
which gives (\ref{delta_gen_eins}), (\ref{Z_eins_gen}), and also the equations
for $T_c$ (\ref{lin_delta_einst}), (\ref{Z_lin_einst}) with simple replacement 
of $\Omega^2_0$ by $\langle\Omega^2\rangle$. In this sense the general structure
``Einstein'' Eliashberg equations conserves also for the case of general phonon 
spectrum. This approximation with identification of $\Omega^2_0$ and
$\langle\Omega^2\rangle$ is always assumed in the following.

For the model of phonon spectrum consisting of discreet set of Einstein
phonons:
\begin{eqnarray}
\alpha^2(\omega)F(\omega)=\sum_i\alpha^2(\Omega_i)\delta(\omega-\Omega_i)\nonumber\\
=\sum_i\frac{\lambda_i}{2}\Omega_i\delta(\omega-\Omega_i)
\label{El-Mc-discr}
\end{eqnarray}
In this case from (\ref{av_sq_w}) we simply obtain:
\begin{equation}
\langle\Omega^2\rangle =\frac{1}{\lambda}\sum_i\lambda_i\Omega_i^2
\label{sqfrqav}
\end{equation}
where $\lambda=\sum_i\lambda_i$.

\subsection{Weak and intermediate coupling}

There is a vast literature on solving Eliashberg equations in the weak or intermediate
coupling region $\lambda<$1 \cite{Scal,All,VIK}. Here we only present qualitative results
for $T_c$, dropping unimportant (for our aims) numerical coefficients $\sim$1. 
In a model with Einstein spectrum of phonons and Coulomb potential $\mu^{\star}$=0 
we have \cite{Diagr}:
\begin{equation}
T_c\sim\Omega_0\exp\left(-\frac{1+\lambda}{\lambda}\right)
\label{McMBCS}
\end{equation}
This expression is in fact close to the results of an exact numerical analysis
performed at a time by McMillan \cite{Scal,All,VIK}\footnote{Here we drop some 
numerical coefficients $\sim$1. If we remember a number of (not so well controlled)
approximations made during the derivation of Eliashberg equations, it becomes clear
that we are not loosing much in accuracy here.}

Similar estimates of $T_c$ can be obtained also in the strong antiadiabatic limit, 
considering Eliashberg equations for $\lambda<$1 in the problem with a narrow electron
band of half -- width $D\sim E_F\ll\Omega_0$ \cite{MS_Eli,MS_Elis,MS_Elias}.
Then the appropriately generalized Eliashberg equations give for the same model with
Einstein spectrum:
\begin{equation}
T_c\sim\frac{D}{1+D/\Omega_0}\exp\left(-\frac{1+\tilde\lambda}{\lambda}\right)
\label{TcantiElias}
\end{equation}
where the effective constant $\tilde\lambda$ was defined above in Eqs. (\ref{A6}), 
(\ref{A7}) and (\ref{lamb_D_Mc}). 

Eq. (\ref{TcantiElias}) interpolates between adiabatic and antiadiabatic limits. 
For $D\gg\Omega_0$ it gives (\ref{McMBCS}), while for $D\ll\Omega_0$ it reduces to:
\begin{equation}
T_c\sim D\exp\left(-\frac{1}{\lambda}\right)
\label{Tcstranti}
\end{equation}
i.e. to BCS -- like expression (weak coupling!), where preexponential factor is
determined not by phonon frequency, but by the electron band half -- width
(Fermi energy), which now plays the role of cutoff parameter for divergence in Cooper
channel. This fact was first noted by Gor'kov in Refs. \cite{Gork_1,Gork_2,Gork_3}.

For more general model of phonon spectrum consisting of discrete set of Einstein
phonons (\ref{El-Mc-discr}) these relations are obviously generalized to  \cite{MS_Elis,MS_Elias}:
\begin{equation}
\lambda=2\sum_i\frac{\alpha^2(\Omega_i)}{\Omega_i}\equiv\sum_i\lambda_i
\label{lamb_i}
\end{equation}
\begin{equation}
\tilde\lambda=2\sum_i\frac{\alpha^2(\Omega_i)D}{\Omega_i(\Omega_i+D)}
=\sum_i\lambda_i\frac{D}{\Omega_i+D}
\equiv\sum_i\tilde\lambda_i
\label{lamb_tild}
\end{equation}
and instead of (\ref{TcantiElias}) we have:
\begin{equation}
T_c\sim
\prod_i\left(\frac{D}{1+\frac{D}{\Omega_i}}\right)^{\frac{\lambda_i}{\lambda}}
\exp\left(-\frac{1+\tilde\lambda}{\lambda}\right)
\label{Tc_opt_i}
\end{equation}
In the simplest case of two Einstein phonons with frequencies $\Omega_1$ and $\Omega_2$ 
this gives:
\begin{equation}
T_c\sim
\left(\frac{D}{1+\frac{D}{\Omega_1}}\right)^{\frac{\lambda_1}{\lambda}}
\left(\frac{D}{1+\frac{D}{\Omega_2}}\right)^{\frac{\lambda_2}{\lambda}}
\exp\left(-\frac{1+\tilde\lambda}{\lambda}\right)
\label{Tc_opt_2}
\end{equation}
where $\tilde\lambda=\tilde\lambda_1+\tilde\lambda_2$ and
$\lambda=\lambda_1+\lambda_2$. In case of $\Omega_1\ll D$ (adiabatic phonon),
and $\Omega_2\gg D$ (antiadiabatic phonon) (\ref{Tc_opt_2}) reduces to:
\begin{equation}
T_c\sim (\Omega_1)^{\frac{\lambda_1}{\lambda}}
(D)^{\frac{\lambda_2}{\lambda}}
\exp\left(-\frac{1+\tilde\lambda}{\lambda}\right)
\label{Tc_opt_ad_ant}
\end{equation}

Eq. (\ref{Tc_opt_i}) is easily rewritten as:
\begin{equation}
T_c\sim\langle\Omega_{\ln}\rangle\exp\left(-\frac{1+\tilde\lambda}{\lambda}\right)
\label{Tc_log}
\end{equation}
where we have introduced mean logarithmic frequency $\langle\Omega_{\ln}\rangle$:
\begin{equation}
\ln \langle\Omega_{\ln}\rangle=\ln\prod_i
\left(\frac{D}{1+\frac{D}{\Omega_i}}\right)^{\frac{\lambda_i}{\lambda}}
=\sum_i\frac{\lambda_i}{\lambda}\ln \frac{D}{1+\frac{D}{\Omega_i}}
\label{Omega_ln}
\end{equation}
In the limit of continuous distribution of phonon frequencies the last
expression reduces to:
\begin{equation}
\ln\langle\Omega_{\ln}\rangle = \frac{2}{\lambda}\int\frac{d\omega}{\omega}
\alpha^2(\omega)F(\omega)\ln\frac{D}{1+\frac{D}{\omega}}
\label{Omega_log}
\end{equation}
where $\lambda$ is given by the usual relation (\ref{lambda_Elias_Mc}).

In general case, the preexponential factor in the expression for $T_c$ in
Eliashberg theory for weak or intermediate coupling is always given by mean
logarithmic phonon frequency \cite{VIK}, and Eq. (\ref{Omega_log}) gives
the generalization of the standard expression for this frequency for the case of
electron band of finite width. From Eq. (\ref{Omega_log}) we can easily obtain
the standard expression \cite{VIK} (see also below) for adiabatic limit,  
when $D\to\infty$.

\subsection{Lower bound for $T_c$ in the limit of very strong coupling}

To achieve really high values of $T_c$ the region of strong and very strong
coupling $\lambda>1$ is of the main interest and will be discussed in the
following. The general Eliashberg equations in Matsubara representation,
determining superconducting gap $\Delta(\omega_n)$ at arbitrary temperatures,
are given in (\ref{eq_delta}), (\ref{Zgen}) \cite{Scal,All}.

Limitations on the value of $T_c$ in the limit of very strong coupling are
easily derived analytically. We shall see shortly, that appropriate behavior
follows form the estimate of the {\em lower} bound for $T_c$ \cite{AD}.
Consider the linearized gap equation (\ref{lin_delta}), determining $T_c$:
\begin{eqnarray}
\Delta(\omega_n)Z(\omega_n)=\pi T\sum_{n'}\int_{0}^{\infty}\alpha^2(\omega)
F(\omega)\times\nonumber\\
\times D(\omega_n-\omega_{n'};\omega)\frac{\Delta(\omega_{n'})}{|\omega_{n'}|}
\label{lin_delta_gen}
\end{eqnarray}
where phonon Green's function is defined in (\ref{Dw}). Let us consider the
term with $n=0$. Then, leaving in the sum in Eq. (\ref{Z_lin_einst}) only
the contribution from $n'=0$, we obtain:
\begin{equation}
Z(0)=1+\lambda
\label{Zlamb}
\end{equation}
which after the substitution into Eq. (\ref{lin_delta_gen}) for  $n=0$ just cancels 
the similar (corresponding to $n'=0$) term in the r.h.s. of Eq. \cite{Kres,Kresin_Gut}, 
so that the equation for $\Delta(0)=\Delta(\pi T)$ takes the form:
\begin{equation}
\Delta(0)=\pi T\sum_{n'\neq 0}\int_{0}^{\infty}\alpha^2(\omega)
F(\omega)\frac{2\omega}{(\pi T-\omega_{n'})^2+\omega^2}\frac{\Delta(\omega_{n'})}{|\omega_{n'}|}
\label{lin_delta_0}
\end{equation}
All terms in the r.h.s. here are positive. Let us leave only the contribution 
from $n'=-1$, so that taking into account $\Delta(-1)=\Delta(-\pi T)=\Delta(\pi T)=\Delta(0)$, 
and canceling $\Delta(0)$ in l.h.s and r.h.s. we immediately get the {\em inequality}
\cite{AD}:
\begin{equation}
1>\int_{0}^{\infty}d\omega\frac{2\alpha^2(\omega)F(\omega)\omega}
{(2\pi T)^2+\omega^2}
\label{AD_ineq}
\end{equation}
Actually here $T=T_c$, and this equation gives the {\em lower}  estimate of $T_c$. 
In particular, in the model with Einstein spectrum of phonons
$F(\omega)=\delta(\omega-\Omega_0)$ and this inequality is immediately rewritten as: 
\begin{equation}
1>2\alpha^2(\Omega_0)\frac{\Omega_0}{(2\pi T)^2+\Omega_0^2}=
\lambda\frac{\Omega^2_0}{(2\pi T)^2+\Omega_0^2}
\label{AD_in}
\end{equation}
so that for $T_c$ we get:
\begin{equation}
T_c>\frac{1}{2\pi}\sqrt{\lambda-1}\Omega_0
\label{TcAD}
\end{equation}
which for $\lambda\gg 1$ reduces to:
\begin{equation}
T_c>\frac{1}{2\pi}\sqrt{\lambda}\Omega_0\approx 0.16\sqrt{\lambda}\Omega_0
\label{TcA_D}
\end{equation}
\begin{figure}
\includegraphics[clip=true,width=0.45\textwidth]{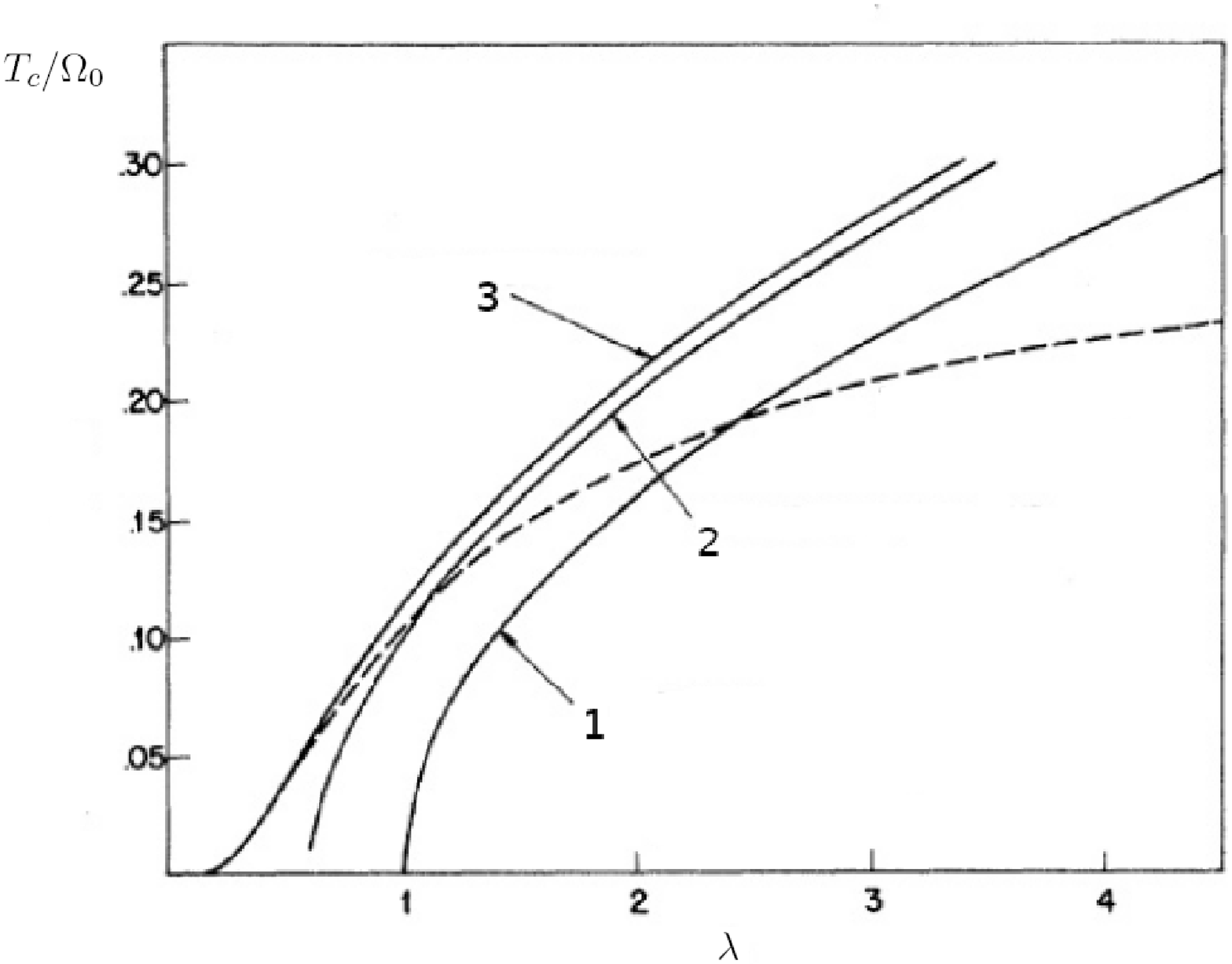}
\caption{Temperature of superconducting transition in Einstein model
of phonon spectrum in units of $T_c/\Omega_0$ as a function of pairing constant
$\lambda$ \cite{AD}: 1 -- lower bound (\ref{TcAD}),
2 -- solution of the system of two linear equations ($n=0,\pm 1$), 
3 -- numerically exact solution of the full system of equations ($n\leq 63$) \cite{AD}.
McMillan expression for $T_c$ \cite{VIK} is shown by dashed line (for the case of
$\mu^{\star}$=0).}
\label{TcADcomp}
\end{figure}

For discrete spectrum of phonons (\ref{El-Mc-discr}) inequality (\ref{AD_ineq})
reduces to:
\begin{equation}
1>\sum_i\lambda_i\frac{\Omega_i^2}{(2\pi T)^2+\Omega_i^2}
\label{ADIneqdiskr}
\end{equation}
which in the limit of very strong coupling for $2\pi T\gg\Omega_i$ immediately
gives the natural generalization of (\ref{TcA_D}):
\begin{equation}
T_c > \frac{1}{2\pi}\sqrt{\lambda\langle\Omega^2\rangle}
\label{Tc_Al_Dy}
\end{equation}
where $\lambda=\sum_i\lambda_i$ and $\langle\Omega^2\rangle$ was defined above in
Eq. (\ref{sqfrqav}).

If we solve 2$\times$2 system of equations and appropriate quadratic equation
for $T_c$, following from Eqs. (\ref{lin_delta_einst}), (\ref{Z_lin_einst}) for 
$n=0;\pm 1$, the constant 0.16 in (\ref{TcA_D}) is replaced by  0.18. The solution
of system with $n\leq$63, performed in Ref. \cite{AD} lead to replacement of
0.18 by 0.182, which practically corresponds to numerically exact solution. 
It is obvious, that even the simplest solution (\ref{TcAD}) is quite sufficient for 
qualitative estimates of $T_c$ in the limit of very strong coupling. 
The general situation is illustrated in Fig. \ref{TcADcomp}. From this figure
it can be seen, in particular, that asymptotic behavior of $T_c$
for $\lambda\gg$1 (\ref{TcA_D}) with coefficient 0.18, rather well approximates
the values of critical temperature already starting from the values 
of $\lambda>$1.5-2.0 (cf. curve 2 in this figure).


Consider now the very strong coupling case in strong antiadiabatic limit,
though realization of such coupling in this limit is rather doubtful, as pairing
constant $\lambda$, defined according to (\ref{Elias_lambda_Om}),  typically 
rapidly drops with the growth of phonon frequency, as it exceeds Fermi energy
\cite{MS_Eli,MS_Elias}.

Consider again the electron band of finite width $2D$ (constant density of states). 
Then in general Eliashberg equations considered above, instead of integral in
infinite limits (\ref{intxi}) we have:
\begin{eqnarray}
&&\int_{-D}^{D}d\xi\frac{1}{\omega_{n'}^2+\xi^2+\Delta^2(\omega_{n'})}=\nonumber\\
&&=\frac{2}{\sqrt{\omega_{n'}^2+\Delta^2(\omega_{n'})}}
arctg\frac{D}{\sqrt{\omega_{n'}^2+\Delta^2(\omega_{n'})}}\to\nonumber\\
&&\to\frac{2}{|\omega_{n'}|} 
arctg\frac{D}{|\omega_{n'}|} \ \mbox{for}\ \Delta(\omega_{n'})\to 0
\label{int_xii}
\end{eqnarray}

Then the linearized Eliashberg equations take the following general form:
\begin{eqnarray}
&&\Delta(\omega_n)Z(\omega_n)=\nonumber\\
&&=T\sum_{n'}\int_{0}^{\infty}d\omega\alpha^2(\omega)F(\omega)
D(\omega_n-\omega_{n'};\omega)\times\nonumber\\
&&\times \frac{2\Delta(\omega_{n'})}{|\omega_{n'}|}
arctg\frac{D}{|\omega_{n'}|}
\label{lin_Delta_gen}
\end{eqnarray}
where
\begin{eqnarray}
&&Z(\omega_n)=1+\frac{T}{\omega_n}\sum_{n'}\int_{0}^{\infty}d\omega
\alpha^2(\omega)F(\omega)\times\nonumber\\
&&\times D(\omega_n-\omega_{n'};\omega) \frac{\omega_{n'}}{|\omega_{n'}|}
arctg\frac{D}{|\omega_{n'}|}
\label{lin_Z_gen}
\end{eqnarray}
Now we directly obtain the equation for $\Delta(0)$:
\begin{eqnarray}
&&\Delta(0)=T\sum_{n'\neq 0}\int_{0}^{\infty}d\omega\alpha^2(\omega)F(\omega)
\frac{2\omega}{(\pi T-\omega_{n'})^2+\omega^2}\times\nonumber\\
&&\times\frac{2\Delta(\omega_{n'})}{|\omega_{n'}|}
arctg\frac{D}{|\omega_{n'}|}
\label{D0_eq}
\end{eqnarray}
Again, taking into account only the contribution of $n'=-1$ in the r.h.s., we immediately 
obtain an inequality:
\begin{equation}
1>\frac{2}{\pi}\int_{0}^{\infty}d\omega\alpha^2(\omega)F(\omega)
\frac{2\omega}{(2\pi T)^2+\omega^2}arctg\frac{D}{\pi T}
\label{AD_inq_gen}
\end{equation}
In Einstein model of phonon spectrum we have $F(\omega)=\delta(\omega-\Omega_0)$,
so that Eq. (\ref{AD_inq_gen}) reduces to:
\begin{equation}
1>\frac{2}{\pi}\lambda arctg\frac{D}{\pi T}\frac{\Omega^2_0}{(2\pi T)^2+\Omega^2_0}
\label{AD_gen_ineq}
\end{equation}
For $D\gg\pi T$ it immediately gives the result of Allen and Dynes:
\begin{equation}
T_c>\frac{1}{2\pi}\sqrt{\lambda-1}\Omega_0\to 0.16\sqrt{\lambda}\Omega_0\ \mbox{for}\ \lambda\gg 1\
\label{T_c_AD}
\end{equation}
For $D\ll\pi T$ Eq. (\ref{AD_gen_ineq}) gives:
\begin{equation}
T>\frac{1}{2\pi}\sqrt{\lambda^*(T)-1}\Omega_0
\label{TcAD-star}
\end{equation}
where
\begin{equation}
\lambda^*(T)=\frac{2D}{\pi^2 T}\lambda
\label{lamb-star}
\end{equation}
so that in strong antiadiabatic limit we have: 
\begin{equation}
T_c>(2\pi^4)^{-1/3}(\lambda D\Omega_0^2)^{1/3}\approx 0.17(\lambda D\Omega_0^2)^{1/3}
\label{Tc-AD-anti}
\end{equation}
From the obvious condition $\lambda^*(T)>0$ we get:
\begin{equation}
T_c<\frac{2}{\pi^2}\lambda D\approx 0.202\lambda D
\label{ineq_T}
\end{equation}
which bounds $T_c$ from above.
\begin{figure}
\includegraphics[clip=true,width=0.45\textwidth]{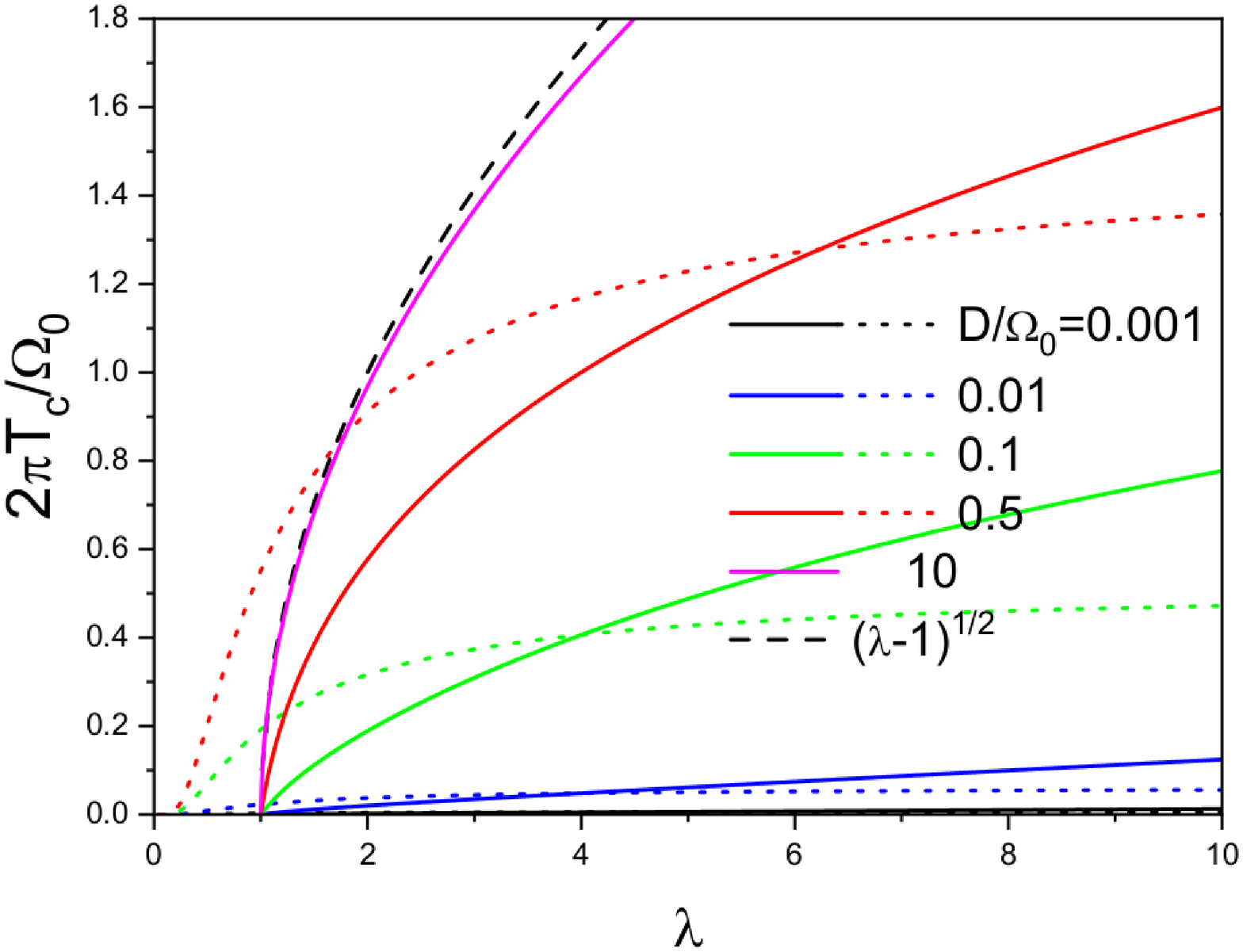}
\caption{Temperature of superconducting transition (lower bound) in 
Einstein model of phonon spectrum in units of $2\pi T_c/\Omega_0$, 
as a function of pairing constant $\lambda$ for different values of inverse 
adiabaticity parameter  $D/\Omega_0$. 
Dashed lines show appropriate dependencies for
$2\pi T_c/\Omega_0$ in the region of weak and intermediate coupling (\ref{TcantiElias}). 
Black dashed line --- Allen -- Dynes estimate (\ref{TcAD}),
valid in adiabatic limit \cite{AD}}.
\label{TcAD_anti_gen}
\end{figure}
\begin{figure}
\includegraphics[clip=true,width=0.45\textwidth]{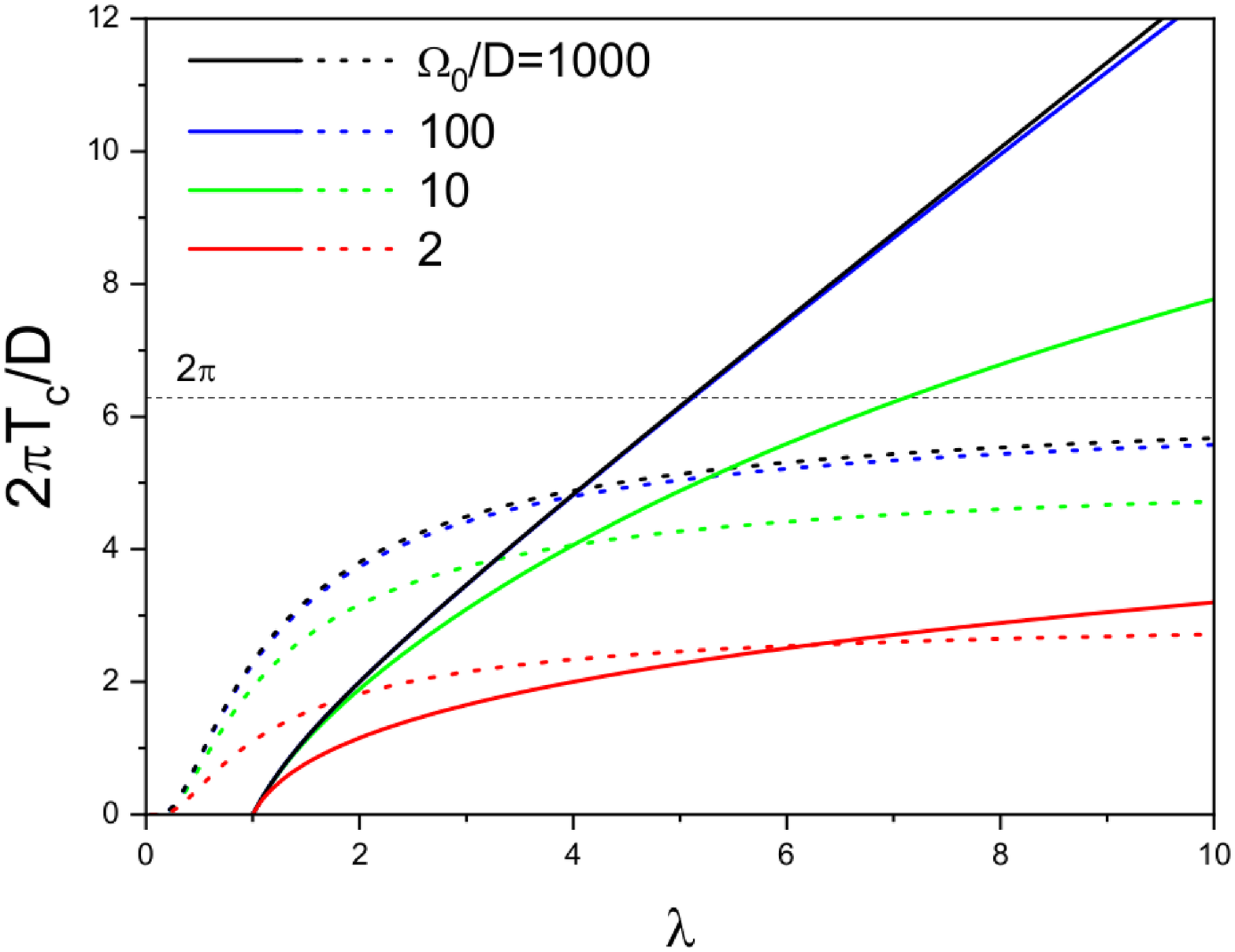}
\caption{Temperature of superconducting transition (lower bound) in Einstein model
of phonon spectrum in units of $2\pi T_c/D$, as a function of pairing constant
$\lambda$ for different values of adiabaticity parameter $\Omega_0/D$.
Dashed lines show appropriate dependencies for $2\pi T_c/D$ in weak and intermediate
coupling region (\ref{TcantiElias}).} 
\label{TcAD_anti_gen_D}
\end{figure}

Thus we have to satisfy an inequality:
\begin{equation}
(2\pi^4)^{-1/3}(\lambda D\Omega_0^2)^{1/3}<T_c<\frac{2}{\pi^2}\lambda D
\label{doubl_ineq}
\end{equation}
which reduces to the condition:
\begin{equation}
\Omega_0<\frac{4}{\pi}\lambda D\approx 1.27\lambda D\ \mbox{or}\
\frac{D}{\Omega_0}>\frac{0.78}{\lambda}
\label{ineq_D_lam}
\end{equation}
so that for self -- consistency of our analysis we have to satisfy the
condition:
\begin{equation}
\lambda\gg\frac{\Omega_0}{D}\gg 1
\label{lamb_big}
\end{equation}
where the last equality corresponds to the limit of strong antiadiabaticity.
Correspondingly, all the previous estimates become invalid for
$\lambda\sim 1$ and can describe only the limit of very strong coupling.

In Fig. \ref{TcAD_anti_gen} and Fig. \ref{TcAD_anti_gen_D} we show the results
of numerical calculations of $T_c$ boundaries, following from the solution of
(\ref{AD_gen_ineq}), as compared to the values of $T_c$ in the region of weak 
and intermediate coupling (\ref{TcantiElias}), for different values of
adiabaticity parameter $\Omega_0/D$. 
It is clear that in the vicinity of crossing dashed and continuous lines in
graphs, shown in Fig. \ref{TcAD_anti_gen} and Fig. \ref{TcAD_anti_gen_D}, 
we actually have a smooth crossover from $T_c$ behavior in the region of weak and
intermediate coupling to its asymptotic behavior in very strong coupling region of
$\lambda\gg 1$. From Fig. \ref{TcAD_anti_gen_D} it is seen that the boundary (\ref{ineq_T}) 
is practically achieved in the region of large values of $\lambda$ and $\Omega_0/D$.
From these figures we can also see that simple rising of phonon frequency
and transition to antiadiabatic limit do not lead, in general, to an increase of
$T_c$ as compared with adiabatic case.


\section{Maximal $T_c$?}

The problem of maximal possible temperature of superconducting transition has
arisen immediately after the creation of BCS theory. It was studied in numerous
papers with sometimes contradictory results. Among these papers was the notorious
paper by Cohen and Anderson \cite{CA}, where rather elegant arguments were given,
seemingly quite convincing, that characteristic scale of $T_c$ values due to
electron -- phonon, or any other similar mechanism, based on exchange of Bose --
like excitations in metals, can be of the order of about 10--30 K only.
This paper was immediately seriously criticized in Refs. \cite{MaxKh,DMK}, 
with conclusion that in reality there are no such limitations.
Even more, the analysis performed in these papers has shown that Ref. \cite{CA}
is just erroneous. However, the point of view expressed in Cohen -- Anderson paper
became popular in physics community (Anderson himself till the end of his life 
adhered to the view expressed in Ref. \cite{CA}), so that at the time of discovery of 
high -- temperature superconductivity in cuprates (1986 -- 1987), almost total belief
was that the ``usual'' electron -- phonon mechanism does not allow values of
$T_c$ higher, that 30-40 K. Because of this after the discovery of superconductivity
in cuprates the ``great race'' has started for new theoretical models and mechanisms
of superconductivity, which may explain the high values of $T_c$. The problems of
superconductivity in cuprates are outside the scope of this work. Most probably, 
in cuprates really dominates some kind of non -- phonon pairing mechanism
(due to antiferromagnetic fluctuations). But most important result of discovery of
record values of $T_c$ in hydrides under high pressures, in our opinion, is the
final (and experimental!) rebuttal of the point of view expressed in Ref. \cite{CA}.

Thus, the problem of maximal value of $T_c$, which may be achieved due to
electron -- phonon mechanism of Cooper pairing is most important as ever.
Below we try to discuss this problem once again within standard approach,
based on Eliashberg equations, as most successful theory, describing 
superconductivity in the system of electrons and phonons in metals.

There is a vast literature on numerical solution of Eliashberg equations for
different temperatures and different models of phonon spectrum \cite{VIK,All}.

A number of analytic expressions for $T_c$ were proposed by different authors, to 
approximate the results of numerical computations. As an example we quote here
the popular interpolation formula for $T_c$ due to Allen and Dynes \cite{All}, 
which is appropriate for rather wide interval of values of dimensionless
coupling constant of electron -- phonon interaction $\lambda$, including the 
strong coupling region of $\lambda>1$:
\begin{equation}
T_{c}=\frac{f_{1}f_{2}}{1.20}\langle\Omega_{\ln}\rangle
\exp\left\{-\frac{1.04(1+\lambda)}{\lambda-\mu^{\star}(1+0.62\lambda)}\right\}
\label{ADTc}
\end{equation}
where
\begin{eqnarray}
&&f_{1}=[1+(\lambda/\Lambda_{1})^{3/2}]^{1/3};\
f_{2}=1+\frac{[\langle\Omega^{2}\rangle^{1/2}/\langle\Omega_{\ln}\rangle-1]\lambda^{2}}{\lambda^{2}
+\Lambda_{2}^{2}} \nonumber \\
&&\Lambda_{1}=2.46(1+3.8\mu^{\star});\
\Lambda_{2}=1.82(1+6.3\mu^{\star})\frac{\langle\Omega^{2}\rangle^{1/2}}
{\langle\Omega_{\ln}\rangle}\nonumber\\
\label{ffll}
\end{eqnarray}
Here $\langle\Omega_{\ln}\rangle$ is the mean logarithmic frequency of phonons: 
\begin{eqnarray}
&&\ln\langle\Omega_{\ln}\rangle = \frac{2}{\lambda}\int_{0}^{\infty}\frac{d\omega}{\omega}
\alpha^2(\omega)F(\omega)\ln\omega =\nonumber\\
&&=\frac{\int_{0}^{\infty} \frac{d\omega}{\omega}\ln\omega
\alpha^2(\omega)F(\omega)}{\int_{0}^{\infty} \frac{d\omega}{\omega}
\alpha^2(\omega)F(\omega)}
\label{Omega_logg}
\end{eqnarray}
$\langle\Omega^{2}\rangle$ is the average (over the phonon spectrum) square of frequency,
defined in Eq. (\ref{av_sq_w}):
\begin{equation}
\langle\Omega^2\rangle = \frac{2}{\lambda}\int_{0}^{\infty}d\omega
\alpha^2(\omega)F(\omega)\omega=\frac{\int_{0}^{\infty} d\omega\omega
\alpha^2(\omega)F(\omega)}{\int_{0}^{\infty} \frac{d\omega}{\omega}
\alpha^2(\omega)F(\omega)}
\label{Omega_sqr}
\end{equation}
The Coulomb pseudopotential $\mu^{\star}$ determines repulsion between electrons
within Cooper pair. According to most calculations \cite{VIK,All} its values are small 
and belong to the interval 0.1 -- 0.15.

In the limit of very strong coupling $\lambda>10$ this gives the expression for $T_c$, 
which was. in fact. obtained above from simple inequality (\ref{TcAD}):
\begin{equation}
T_c\approx 0.18\sqrt{\lambda\langle\Omega^2\rangle}
\label{AD_asymp}
\end{equation}
It may seem now, that limitations on the values of $T_c$ are simply absent,
so that in the limit of very strong coupling very high $T_c$ can be obtained from 
electron -- phonon mechanism. The only more or less obvious limit is related to
the limits of adiabatic approximation, which is usually considered to be the
cornerstone of Eliashberg theory. However, we have seen above. that similar results
can be obtained also in the strong antiadiabatic limit
(cf. estimated for  $T_c$ given in Eqs. (\ref{Tc-AD-anti}), (\ref{ineq_T})).

In the model with Einstein spectrum of phonons we simply have:
$\langle\Omega_{\ln}\rangle=\langle\Omega^{2}\rangle^{1/2}=\Omega$,
where $\Omega$ is assumed to be the renormalized phonon frequency.
Then (\ref{AD_asymp}) reduces to:
\begin{equation}
T_c=0.18\sqrt{\lambda}\Omega
\label{ADyn}
\end{equation}
so that seemingly for $\lambda\gg 1$ we can, in principle, obtain even $T_c>\Omega$. 
However, if we remember the renormalization of phonon spectrum and take into account
Eq. (\ref{phn_spc_dressed_eins}), we immediately obtain from Eq. (\ref{ADyn}):
\begin{equation}
T_c=0.18\sqrt{\lambda}\Omega=0.18\Omega_0\sqrt\frac{\lambda}{1+2\lambda}
\label{ADynes}
\end{equation}
which in the limit of $\lambda\gg 1$ tends to the value $T_{c}^{max}\approx 0.13\Omega_0$,
because of significant softening of phonon spectrum.
At the same time, as noted above, the physical meaning of ``bare'' frequency
$\Omega_0$ in a metal is poorly defined, and in particular it can not be determined from
experiments. Correspondingly, the estimate of Eq. (\ref{ADynes}) somehow hangs in the air.

However, this analysis is valid only under the condition of rigid fixation of
all relation between ``bare'' and ``dressed'' phonon spectra. If we ``forget''
about ``bare'' spectrum of phonons and consider parameters $\Omega$ and
$\lambda$ {\em independent}, we can obtain from Eq. (\ref{ADyn}) very high values
of $T_c$. A certain , though rather artificial model, leading precisely to this
kind of behavior was recently introduced in Ref. \cite{KivBerg}. It considers the
interaction of $N$--component electrons with  $N\times N$--component system of
Einstein phonons in the limit of $N\to\infty$. It was shown that in this model the
renormalization of phonon spectrum due to interaction with conduction electrons
is suppressed, so that in the limit of very strong coupling with
1$\ll\lambda\ll N$ we always have Allen -- Dynes estimate (\ref{ADyn}) with 
$\Omega=\Omega_0$.

However, the problem her is, that in real situation we never can consider $\Omega$ 
and $\lambda$ as independent parameters simply because of the general relations
(\ref{lambda_Elias_Mc}) and (\ref{av_sq_w}), which express $\lambda$ and 
$\langle\Omega^2\rangle$ via integrals of Eliashberg -- McMillan function
$\alpha^2(\omega)F(\omega)$. In fact, we may rewrite the expression for $T_c$ in
the region of very strong coupling as:
\begin{equation}
T_c=0.18\sqrt{\lambda\langle\Omega^2\rangle}=0.25\left[\int_{0}^{\infty}d\omega
\alpha^2(\omega)F(\omega)\omega\right]^{1/2}
\label{TcaFw}
\end{equation}
in adiabatic case and, correspondingly
\begin{eqnarray}
&&T_c=(2\pi^4)^{-1/3}(\lambda D\langle\Omega^2\rangle)^{1/3}=\nonumber\\
&&=(2\pi)^{-1/3}\left[2D\int_{0}^{\infty}d\omega\alpha^2(\omega)F(\omega)\omega\right]^{1/3}
\label{TcaFwanti}
\end{eqnarray}
in antiadiabatic limit. We see, that these expressions for $T_c$ are completely determined
by integrals of $\alpha^2(\omega)F(\omega)$. 

In famous Ref. \cite{Leav} a simple inequality for $T_c$ was proposed, limiting its value by
the square $A$ under $\alpha^2(\omega)F(\omega)$:
\begin{equation}
T_c\leq 0.2309\int_{0}^{\infty}d\omega\alpha^2(\omega)F(\omega)\equiv
0.2309 A
\label{Leavns}
\end{equation}
For the case of Einstein spectrum of phonons, taking into account Eq. (\ref{lambda_Elias_Mc_opt}),
this inequality can be rewritten as:
\begin{equation}
T_c\leq 0.115\lambda\Omega_0
\label{Leavs}
\end{equation}
This inequality is relatively often used in calculations.

The limitation given by Eq. (\ref{ineq_T}) obtained above in antiadiabatic limit is
essentially quite similar to Eq. (\ref{Leavs}), with replacement $\Omega_0\to 2D$, 
which is quite natural in antiadiabatic limit.

Connection of $\lambda$ and $\langle\Omega^2\rangle$ is markedly expressed in McMillan
formula (\ref{McM_form}) for $\lambda$. If we use this expression in (\ref{AD_asymp}), 
we immediately obtain:
\begin{equation}
T_c=0.18\sqrt{\frac{N(0)\langle I^2\rangle}{M}}
\label{AD_McM}
\end{equation}
where
\begin{eqnarray}
&&\langle I^2\rangle=\frac{1}{[N(0)]^2}\sum_{\bf p}\sum_{\bf p'}
\left|I({\bf p-p'})\right|^2\delta(\varepsilon_{\bf p})
\delta(\varepsilon_{\bf p'})=\nonumber\\
&&=\frac{1}{[N(0)]^2}\sum_{\bf p}\sum_{\bf p'}
\left|\langle{\bf p}|
\nabla V_{ei}({\bf r})|{\bf p'}\rangle\right|^2)\delta(\varepsilon_{\bf p})
\delta(\varepsilon_{\bf p'})=\nonumber\\
&&=\langle |\langle {\bf p}|\nabla V_{ei}({\bf r})|{\bf p'}\rangle|^2\rangle_{FS}
\label{I2}
\end{eqnarray}
so that both $\lambda$ and $\langle\Omega^2\rangle$ just drop out from expression for
$T_c$, which is expressed now simply via the averaged over Fermi surface matrix
element of the gradient of electron -- ion potential, ion mass and electron density of
states at the Fermi level. This expression is convenient for ``first principle'' calculations,
where it is often used, but it does not contain illustrative physical parameters in terms
of which we usually treat $T_c$.

\begin{figure}
\includegraphics[clip=true,width=0.45\textwidth]{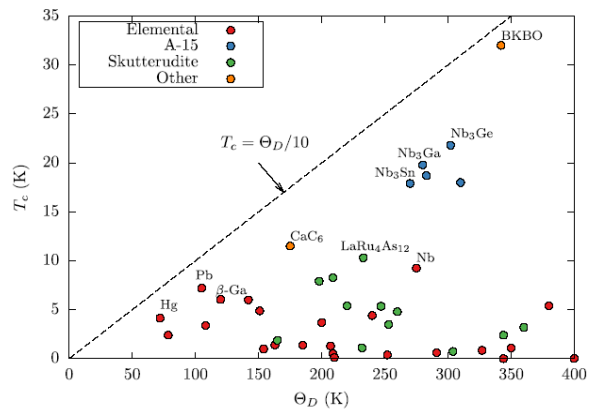}
\caption{Experimental values of the temperature of superconducting transition 
for conventional superconductors dependence on their Debye temperature $\Theta_D$ \cite{Kiv}.}
\label{MaxTcKiv}
\end{figure}

In Ref. \cite{Kiv} a new semiempirical limit for $T_c$ was proposed for conventional 
(electron -- phonon) semiconductors, which is written in a very simple form:
\begin{equation}
k_BT_c\leq A_{max}\Theta_D=A_{max}\hbar\Omega_D
\label{Kiv_Scal}
\end{equation}
where $A_{max}\approx$ 0.10, and $\Theta_D=\hbar\Omega_D$ is Debye temperature, 
which may be determined e.g. from standard measurements of specific heat.
This inequality obviously correlated with $T^{max}_{c}$, obtained above in the
limit of $\lambda\to\infty$ in Eq. (\ref{ADynes}), if we identify $\Omega_0$ with $\Omega_D$. 
It is seen from Fig. \ref{MaxTcKiv} this limitation is satisfied for most of conventional
superconductors \cite{Kiv}. Below we shall see, that is apparently not so in
superhydrides.

\section{Superhydrides and Eliashberg theory}

In this Section we shall briefly discuss some results of application of Eliashberg
theory to $T_c$ calculations in hydrides under high pressures. Our presentation here
will be very short, much more details may be found, for example, in  reviews
\cite{Grk-Krs,ErPR,Ash} and in original papers, some of these will be quoted below.

Eliashberg equations were widely used for calculations of $T_c$ in hydrides.
Actually, both crystal structure of H$_3$S under high pressures and high values of
$T_c\sim$200K were {\em predicted} in Ref. \cite{Duan}. For the structure $Im-3m$ 
under pressure of 200GPa, they obtained the value of pairing constant
$\lambda\approx$ 2.2 and mean logarithmic frequency of phonons
$\Omega_{ln}\approx$ 1335K, so that for $T_c$ calculated from Allen -- Dynes
expression (\ref{ADTc}) with Coulomb pseudopotential values $\mu^{\star}=$0.1--0.13 
the values of $T_c=$191K--204K were obtained. These results were found to be in quite
satisfactory agreement with experiment \cite{H3S}. The achievement of room
temperature values of $T_c$ in  C-S-H system \cite{RT} was reasonably explained in a
recent paper \cite{Ge}, where it was shown that hole doping of $Im-3m$ structure
of H$_3$S by introduction of carbon shifts Fermi level to a maximum of Van -- Hove
singularity in the density of states and certain softening of phonon spectrum.
Combined, all these lead to the growth of $\lambda$ up to the value of 2.4, which is
sufficient, in principle,  to explain the values of $T_c\approx$288K.

\begin{table*}

\centering
\footnotesize
\caption{Calculated values of $T_c$  for La-H and Y-H compounds, obtained from numerical
solution of Eliashberg equations \cite{Ash} compared with its boundary values.}
\label{tab:table_superc}

\begin{tabular}
{@{\hspace{0.3cm}}l@{\hspace{0.3cm}}l@{\hspace{0.3cm}}l@{\hspace{0.3cm}}l@{\hspace{0.3cm}}l@{\hspace{0.3cm}}l@{\hspace{0.3cm}}l@{\hspace{0.3cm}}l@{\hspace{0.3cm}}}

    \hline
    \hline
    Compound    & Pressure (GPa)   &  $\lambda$ &  $\Omega_{\ln}$(K) & $T_c(\mu^{\star}=0.1)$(K) & $T_c$($\mu^{\star}=0.13)$(K) & $\frac{1}{2\pi}\sqrt{\lambda-1}\Omega$ & 0.18$\sqrt{\lambda}\Omega$ \\ [0.15cm]
   \hline
    LaH$_{10}$      & 210 &  3.41    &  848   & 286   & 274 & 209  & 282  \\
    LaH$_{10}$      & 250 &  2.29    &  1253  & 274   & 257 & 226  & 341  \\
    LaH$_{10}$	    & 300 &  1.78    &  1488  & 254   & 241 & 209  & 357  \\
    YH$_{10}$       & 250 &  2.58    &  1282  & 326   & 305 & 256  & 370  \\
    YH$_{10}$	    & 300 &  2.06    &  1511  & 308   & 286 & 247  & 390  \\
    \hline
\end{tabular}
\end{table*}

In Table \ref{tab:table_superc} we show the calculated parameters of several hydrides
of rare -- earth elements from Ref. \cite{Ash}, for which the record values of $T_c$
were predicted. In last two columns of the Table we give the boundaries for $T_c$, 
calculated from inequality (\ref{TcAD}) and from asymptotic expression of Allen and 
Dynes (\ref{ADyn}), under the simplest assumption of
$\langle\Omega_{\ln}\rangle=\langle\Omega^{2}\rangle^{1/2}=\Omega$.
We can see that these values are close enough to those obtained from more detailed
calculations of Ref. \cite{Ash} and determine in fact the lower and upper bounds 
for $T_c$. This clearly shows that the systems with highest values of $T_c$ achieved
are practically already in the very strong coupling region of Eliashberg theory.

In a recent paper \cite{Pick} the extensive calculations of $T_c$ were performed
for practically all possible binary compounds of hydrogen with other elements of
periodic system for the values of external pressure 100, 200 and 300 GPa
(for which the stable crystal structures were also determined).
As many as 36 new systems were discovered for which $T_c$ may exceed 100K, and in
18 cases $T_c$ exceeded 200K. In particular, for NaH$_6$ system the values
of $T_c=$248-279K were obtained, and for CaH$_6$ --- $T_c=$216-253K, already for
pressured of 100 GPa.
The results of this paper clearly show, that the highest possible values of $T_c$ 
are achieved in the region of very strong coupling, up to the values of $\lambda=$5.81 
in NaH$_6$ (under 100 GPa).

Summarizing we may say, that the record values of $T_c$ in superhydrides are achieved
for typical values of $\lambda=$ 2--3.5 (or even more) and for characteristic phonon 
frequencies from 1000K to 2000K. We must also note, that the upper bound expressed by
Eq. (\ref{Kiv_Scal}) is already quite surpassed in some of the known superhydrides.

\section{Conclusions}

We consciously presented all problems related to derivation and use of Eliashberg
equations on sufficiently elementary level, trying to stress all approximations and
simplifications. 

Eliashberg theory remains the main theory, which completely explains the values
of the critical temperature in superconductors with electron -- phonon mechanism
of pairing. This theory is also applicable in the region of strong electron -- phonon
coupling, limited only by the applicability of adiabatic approximation, based on Migdal
theorem, which is valid in the vast majority of metals, including the new superhydrides
with record values of $T_c$. The values of (renormalized, physical) pairing coupling
constant $\lambda$ can surely exceed unity until the system possess the metallic 
ground state. This is not so in the vicinity of phase transition to a new ground
state like charge density wave (CDW) or Bose -- condensate of bipolarons.

More so, Eliashberg theory is qualitatively applicable also in strong antiadiabatic
limit. Simple interpolation expressions for $T_c$ can be constructed, connecting adiabatic
and antiadiabatic regions, The strong antiadiabatic limit may be of importance in
exotic enough systems with very narrow electronic bands and(or) anomalously small
values of Fermi energy (like monolayers of FeSe, SrTiO$_3$ and, probably, some hydrides).

Unfortunately this theory does not produce a simple expression for maximal values
of $T_c$ in terms of experimentally measurable (or calculated) parameters like
characteristic (average) values of phonon frequencies and pairing coupling constant.
Formally, such limit is just absent, if we consider these parameters as independent. 
However, if take into account their interdependence, the maximal values of $T_c$ 
are in fact determined by some ``game'' of atomic constants.
However, all new data on superhydrides strongly indicate that all these systems are 
very close to the strong coupling region coupling of Eliashberg theory, which means
that maximal values of $T_c$ for ``usual'' metals are already achieved. The only hope
probably remains only for metallic hydrogen \cite{Max}.

This work was partially supported by RFBR grant No. 20-02-00011.


\end{document}